\documentclass[preprint, 12pt]{elsarticle}

\usepackage{amsmath}
\usepackage{subfigure}
\usepackage{subfigmat}
\usepackage{multirow}
\usepackage{url}
\usepackage{tabularx}

\journal{Transportation Research Part B}

\begin{document}

\begin{frontmatter}

\title{Numerical Analysis of Gate Conflict Duration and Passenger Transit Time in Airport}

\author[gt]{Sang Hyun Kim\corref{cor}}
\ead{sanghyun.kim@gatech.edu}

\author[gt]{Eric Feron}
\ead{feron@gatech.edu}

\cortext[cor]{Corresponding author. Tel.: +1 404-894-3000}
\address[gt]{School of Aerospace Engineering, Georgia Institute of Technology, Atlanta, GA, 30332, USA}

\begin{abstract}
Robustness is as important as efficiency in air transportation. All components in the air traffic system are connected to form an interactive network. So, a disturbance that occurs in one component, for example, a severe delay at an airport, can influence the entire network. Delays are easily propagated between flights through gates, but the propagation can be reduced if gate assignments are robust against stochastic delays. In this paper, we analyze gate delays and suggest an approach that involves assigning gates while making them robust against stochastic delays. We extract an example flight schedule from data source and generate schedules with increased traffic to analyze how the compact flight schedules impact the robustness of gate assignment. Simulation results show that our approach improves the robustness of gate assignment. Particularly, the robust gate assignment reduces average duration of gate conflicts by 96.3\% and the number of gate conflicts by 96.7\% compared to the baseline assignment. However, the robust gate assignment results in longer transit time for passengers, and a trade-off between the robustness of gate assignment and passenger transit time is presented.
\end{abstract}

\begin{keyword}
	Robust Gate Assignment, Gate Conflict, Passenger Transit, Tabu Search
\end{keyword}

\end{frontmatter}

\section{Introduction}
\subsection{Flight Delays and the Robustness of Gate Assignment}
Severe weather conditions, unanticipated system errors, or incidents at an airport cause flight delays, which are propagated to other airports because the entire air traffic network is connected. From time to time, flight delays accumulate because of the network effect. The network effect comes from airports, or more specifically gates. A single aircraft is assigned to a series of flight legs on a daily basis; a flight leg ends and the next flight leg begins at an airport gate. Therefore, if a flight leg is delayed for some reason, the following flight leg (tail-connected flight) is likely to be delayed and the gate occupied by the aircraft will be released later than the scheduled time. Since the gate is assigned to other aircraft subsequently, a delay of an aircraft may cause serial delays of the following aircraft that are assigned to the same gate. Such a propagation of delay results from the fact that numerous aircraft use airport gates and their gate schedules are dense. For instance, one aircraft is assigned to flight number DL0812, which flies from West Palm Beach, FL to Atlanta, GA. After the flight, the same aircraft is given to another flight number DL1490, which departs from Atlanta, GA to Minneapolis, MN \cite{bts, flightstats}. If the flight incoming to Atlanta, DL0812, arrives late, then the tail flight, DL1490, is also going to depart late. 

\begin{figure}[htb]
 \centering
 \includegraphics[width=0.7\textwidth]{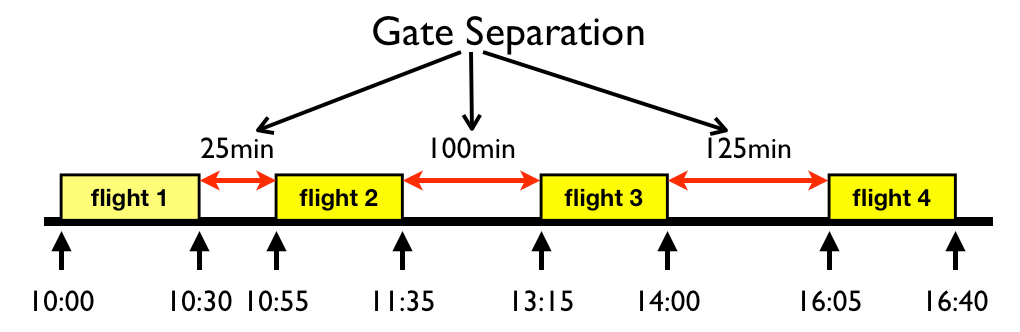}
 \caption{Gate separations.}
 \label{f:separation}
\end{figure}

As discussed above, delays can be propagated through airport gates and the propagated delays are harmful to the efficiency of the air traffic network. A proper gate assignment can help reduce the propagation of delays by absorbing some portion of delays in the time gap between gate schedules. Therefore, robustness becomes an important issue of gate assignment problem because flight delays are uncertain and hard to estimate precisely. Robust gate assignment is thought to be the type of assignment that is insensitive to variations in flight schedules \cite{bolat1999aaf, bolat2000ppr} and that minimizes the number of gate-reassigned aircraft \cite{lim2005rag}. In order to achieve robust gate assignment, many researchers have attempted various methods. For example, buffer times are used to absorb stochastic flight delays to some extent \cite{hassounah1993demand, bolat1999aaf, yan1998network}. Buffer time is the minimally required amount of gate separation, which is the term for the time gap between two consecutive gate schedules. An example of a series of gate schedules and corresponding gate separations are shown in Fig.~\ref{f:separation}.

\subsection{Gate Conflict due to Flight Delays}
If an aircraft is severely delayed and departs later than the scheduled time, the aircraft occupies the assigned gate longer, and the next aircraft that is assigned to the gate may not access the gate on time. Then, the gate assignment is disturbed, and such a situation is called gate conflict. If the previous aircraft is not delayed severely or the amount of departure delay is less than the gate separation between the delayed aircraft and the following aircraft, the gate assignment is not disturbed, and the next aircraft can access the assigned gate without any additional delay. This shows that gate separations influence the probability of gate conflict \cite{kim2011rga} and the punctuality of ramp operations as a result \cite{kim2012iga}. However, if buffer time is too large, the utilization of gates is reduced. To address this issue, there is some research on designing buffer time. Yan et al. developed a simulation framework to analyze stochastic delays and designed buffer times that vary with traffic densities at an airport \cite{yan2002sfe}. They evaluated the trade-off between maximizing gate utilization and maximizing the robustness of gate assignments. Long buffer time makes gate assignments robust against disturbances such as stochastic delays, but is unfavorable for efficient utilization of resources. On the other hand, short buffer time is advantageous for increasing gate usage, but gate assignments become sensitive to small changes in flight schedules.

\subsection{Gate Assignment Problem}
The objectives of traditional gate assignment are minimizing passengers' walking distance \cite{mangoubi1985oga,xu2001aga}, passengers' transferring distance \cite{haghani1998oga}, the number of flights that are assigned to a ramp \cite{ding2003aircraft,ding2005oca}, and aircraft congestion on ramps \cite{kim2009agr,kim2012aga}. Studies of robust gate assignment focus on different objectives. For instance, Bolat studied maximizing idle times of gates (gate separations) \cite{bolat2000ppr}. Since the sum of gate occupancy times is constant, he tried to distribute gate separations evenly among aircraft. Lim and Wang defined gate conflict as occurring when two aircraft are assigned to the same gate and two durations of gate occupancy overlap \cite{lim2005rag}. They assumed that the gate separation determines the probability of gate conflict. They selected several candidate functions to estimate the probability of gate conflict and determined that an exponential function provides the most robust assignment compared to an inverse function, a linear function, and a sublinear function. However, their candidate functions were not based on an analysis of flight delays. Yan and Tang considered the interrelationship between gate assignment and stochastic flight delays \cite{yan2007heuristic}. They randomly generated a finite number of scenarios of flight delays and found the most robust gate assignment based on these scenarios and real-time reassignment rules. Recently, \c{S}eker and Noyan proposed stochastic optimization models that minimize the number of gate conflicts, the variance of gate separation, and the number of gate separations that are shorter than buffer time \cite{cseker2012stochastic}. Similar to Yan and Tang \cite{yan2007heuristic}, \c{S}eker and Nayan analyzed a finite number of scenarios with discrete flight delays. Although buffer time can absorb a minor modification of flight schedules, gate reassignment is occasionally required if a huge delay occurs, gates suddenly become out of order and so forth. For example, Gu and Chung developed a genetic algorithm to reassign gates \cite{gu1999genetic}.

In this paper, we analyze the characteristics of gate delays and model gate delays using a probability distribution. Using the delay model, we calculate the expected duration of gate conflicts based on scheduled times and propose a robust gate assignment policy that minimizes the total (expected) duration of gate conflicts. Also, we analyze the impact of robust gate assignment on the transit time of passengers. Hence, the contributions of this paper are to propose a mathematical metric that evaluates the robustness of gate assignment and to provide numerical analysis on the interrelationship between the robustness of gate assignment and passenger transit time in airport.

\section{Gate Delay Analysis}
Most uncertainties of ramp operations come from stochastic flight delays. Mueller and Chatterji analyzed characteristics of departure and arrival delays \cite{mueller2002aaa}. They selected ten major U.S. airports that experience significant delays and developed models that describe the stochastic nature of delays. They used Normal and Poisson distributions to predict departure, en route, and arrival delays. Tu et al. implemented a genetic algorithm to develop a model that captures seasonal and daily trends in departure delays \cite{tu2008estimating}. Indeed, there are significant differences between seasons; for example, in winter, there are many cancellations and severe delays due to icy weather. They also showed that departure delays gradually stack up during the day and decrease at night. Xu et al. studied delay propagations in the National Airspace System (NAS) \cite{xu2005estimation}. They categorized propagating patterns in turn around processes at Chicago O'Hare International Airport and showed how a delay at an origin airport can propagate to a destination airport. Bruinsma et al. investigated departure and arrival delays in European public transportation systems \cite{bruinsma1999unreliability}. They took Gamma, Log-normal, and Weibull distributions as candidate models that describe delays. Although there are many studies on the development of a model that estimates delays, no specific model or any probability distribution is known to completely describe characteristics of delays.

A gate assignment is disturbed when two consecutive gate schedules overlap, that is, when an aircraft departs too late and/or the next aircraft arrives too early. For instance, suppose that two aircraft are assigned to a gate and their gate schedules are from 1:00PM to 2:00PM and from 2:30PM to 3:30PM. If the departure of the first aircraft is delayed until 2:20PM and the next aircraft arrives on time, this gate assignment is not disturbed. However, if the first aircraft is delayed 20 minutes more, then the gate is occupied when the next aircraft arrives. In this case, the duration of the gate conflict is 10 minutes. Longer duration means that the gate assignment is more disturbed and less robust, and the following aircraft will be delayed. Thus, in this study, the robustness of gate assignments is measured as the duration of gate conflicts. Because delays are uncertain, the expected value of the duration is used to evaluate the robustness of gate assignments.

Therefore, the analysis of gate delays is important to achieve robust gate assignment. The FAA has various databases of flight operations, including the Aviation System Performance Metrics (ASPM), which contains airport data and individual flight data. The airport data give capacities and throughput of airports for every 15 minutes, and the individual flight data provide scheduled and actual gate departure and arrival times and so forth. Also, the Bureau of Transportation Statistics (BTS) provides detailed statistics including scheduled departure/arrival times, actual departure/arrival times, delays, and other information. In order to analyze gate delays, this study uses both ASPM data and airline on-time statistics provided by BTS \cite{bts}.

\subsection{Gate Delay Model}
\begin{figure}[htb]
\begin{subfigmatrix}{2}
 \subfigure[Departure delay distribution of NWA at DTW in March 2006 and the fitted PDF.]{\label{f:depdelay}\includegraphics{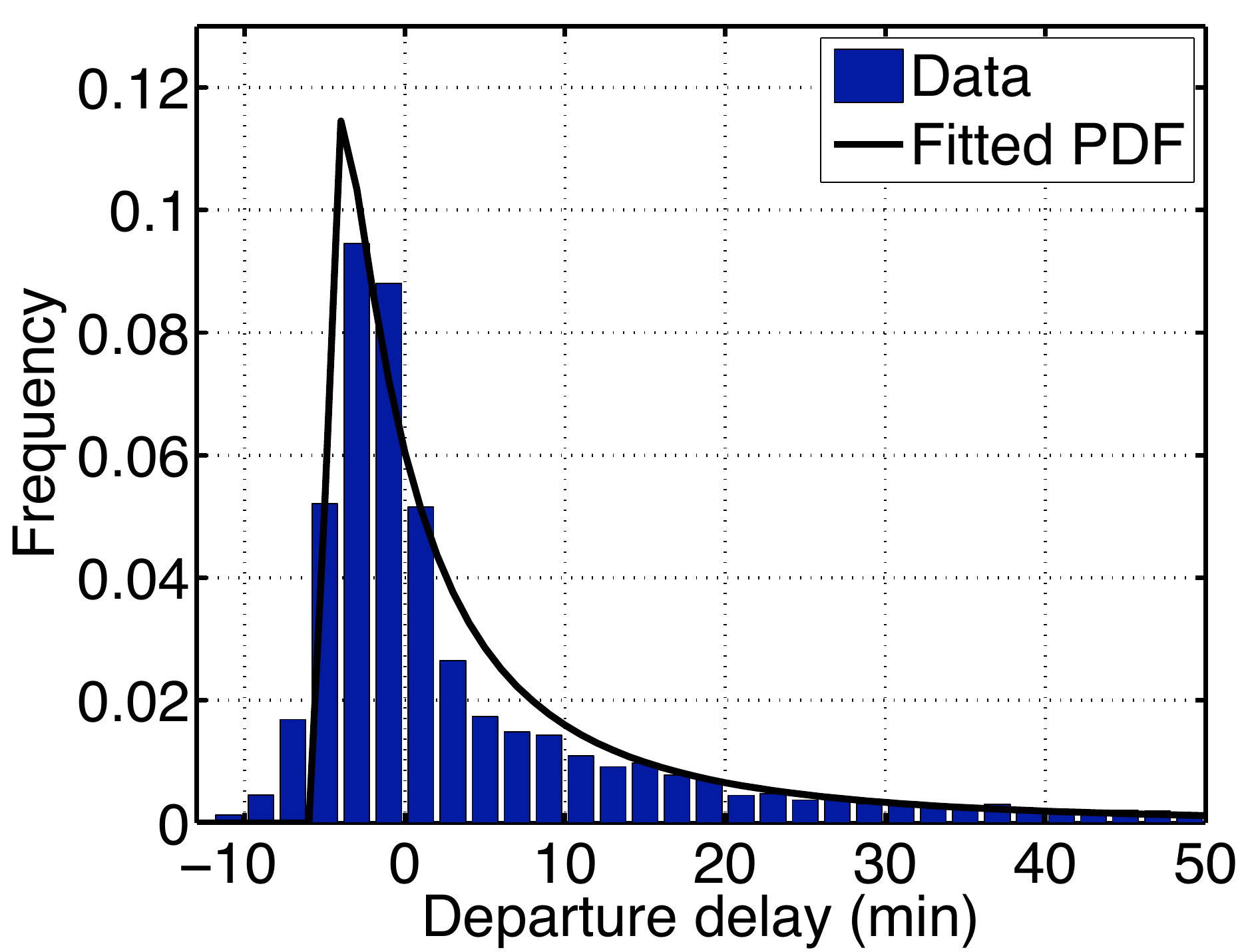}}
 \subfigure[Arrival delay distribution of NWA at DTW in March 2006 and the fitted PDF.]{\label{f:arrdelay}\includegraphics{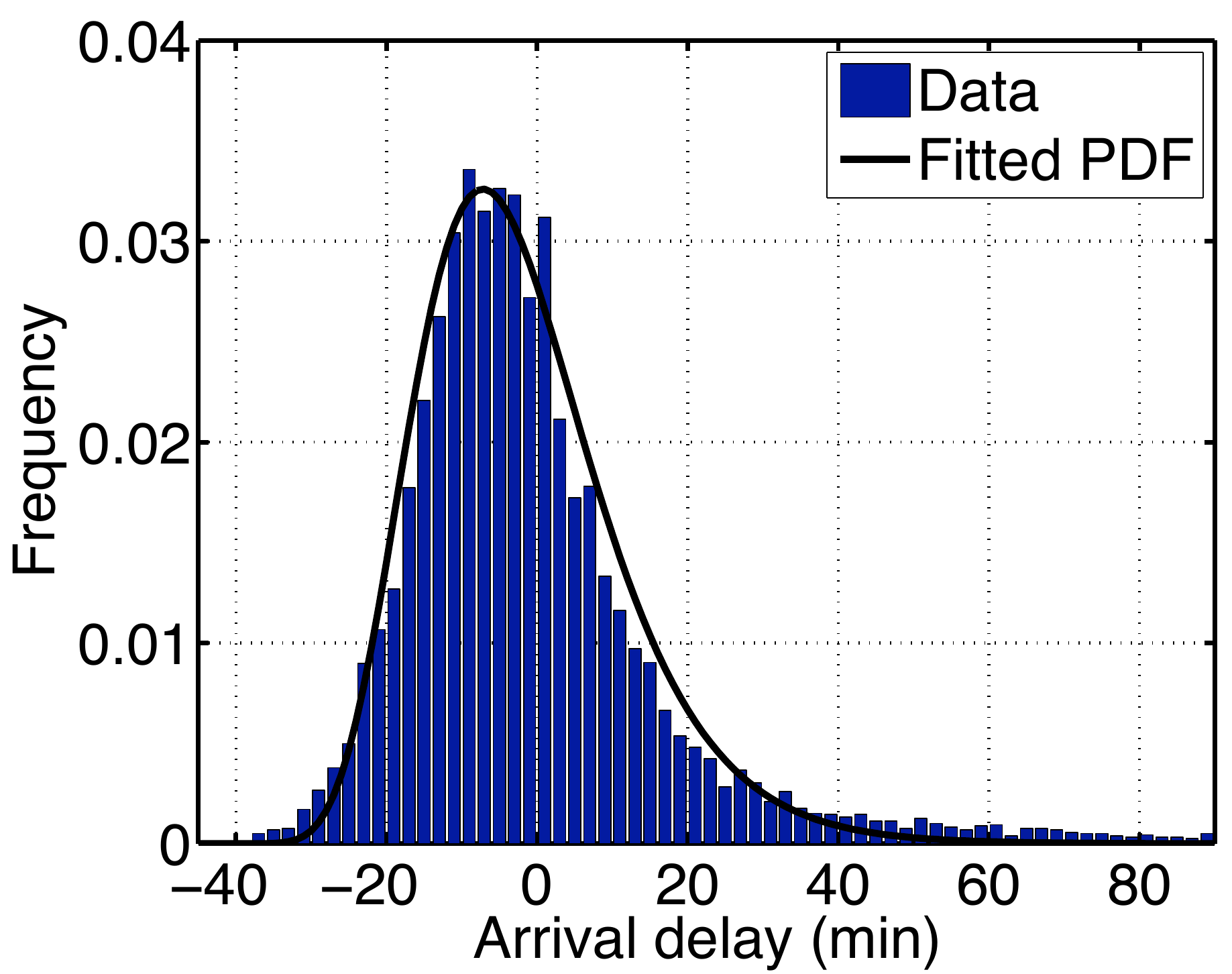}}
\end{subfigmatrix}
 \caption{Delay distributions of NWA at DTW and the fitted PDFs.}
 \label{f:delay}
\end{figure}

As an example, gate delays of Northwest Airlines (NWA) at Detroit Metropolitan Wayne County Airport (DTW) in March 2006 are analyzed using airline on-time statistics provided by BTS \cite{bts}. During that period, there were 7923 departures and 7923 arrivals. The minimum and maximum of departure delays are -15 minutes (15 minutes earlier than scheduled) and 380 minutes, and those of arrival delays are -48 minutes and 1245 minutes. The average and median of departure delays are 8.07 minutes and -1 minute, and those of arrival delays are 2.61 minutes and -4 minutes. The standard deviations of departure and arrival delays are 24.45 minutes and 39.83 minutes, respectively. The statistical characteristics tell that the distributions of departure and arrival delays shift to the left and arrival delays are distributed more sparsely than departure delays. Distributions of departure and arrival delays are shown in Fig.~\ref{f:delay}.

A Log-normal distribution is used to model gate delays. A Log-normal distribution is bounded to nonnegative numbers, but delays can have negative values (early departure or arrival). So, a shift parameter is necessary to capture the characteristic of delays. The probability density function (PDF) of shifted Log-normal distribution is given in Eq.~(\ref{e:lognpdf}), where $\mu$ is the mean of natural logarithm of unshifted delays, $\sigma$ is the standard deviation of natural logarithm of unshifted delays, and $c$ is the shift parameter. The variable $X$ denotes departure delays or arrival delays. Parameters of fitted PDFs for departure and arrival delays of NWA at DTW in March 2006 are given in Table~\ref{t:fit}, and the fitted PDFs are shown in Fig.~\ref{f:delay}.
\begin{align}
	&X \sim \mathbf{Log-}\mathcal{N}(\mu, \sigma) + c, \nonumber \\
	\label{e:lognpdf}
	&f_{X}(X; \mu, \sigma, c) = \frac{1}{(X-c) \sigma \sqrt{2\pi}} e^{-\frac{(\ln (X-c) -\mu)^2}{2 \sigma ^2}}, \forall X>c.
\end{align}

\begin{table}[htb]
	\begin{center}
	\caption{Parameters of the fitted PDFs}
	\label{t:fit}
	\begin{tabularx}{0.5\textwidth}{lXXX}
		\hline
		& $\mu$ & $\sigma$ & $c$ \\ \hline
		Departure & 1.802 & 1.242 & -5.275 \\
		Arrival & 3.812 & 0.2814 & -49 \\
		\hline
	\end{tabularx}
	\end{center}
\end{table}

\begin{figure}[htb]
	\begin{subfigmatrix}{2}
	\subfigure[Departure delay distribution of DAL at ATL in May 2011.]{\label{f:atldepdelay}\includegraphics{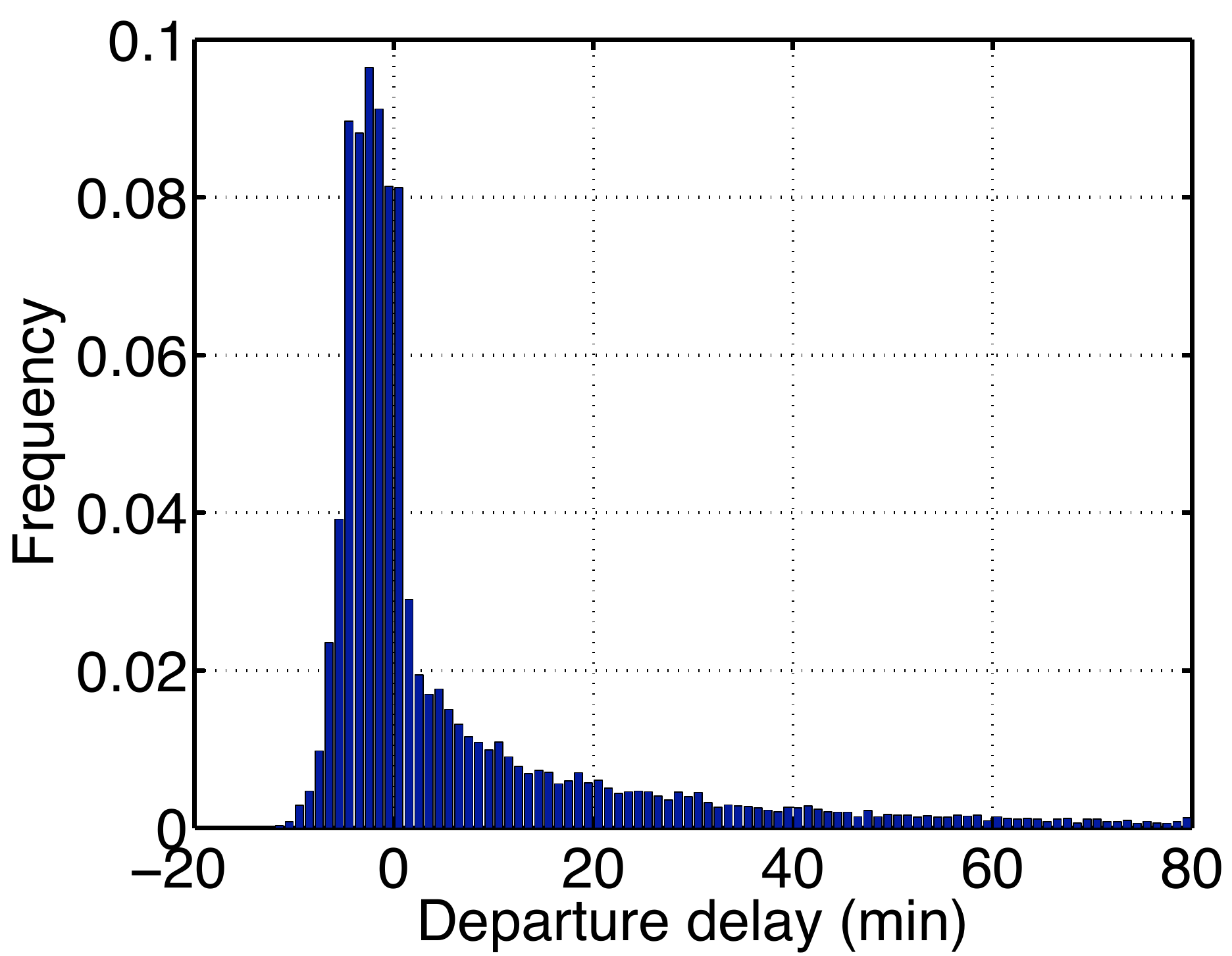}}
	\subfigure[Arrival delay distribution of DAL at ATL in May 2011.]{\label{f:atlarrdelay}\includegraphics{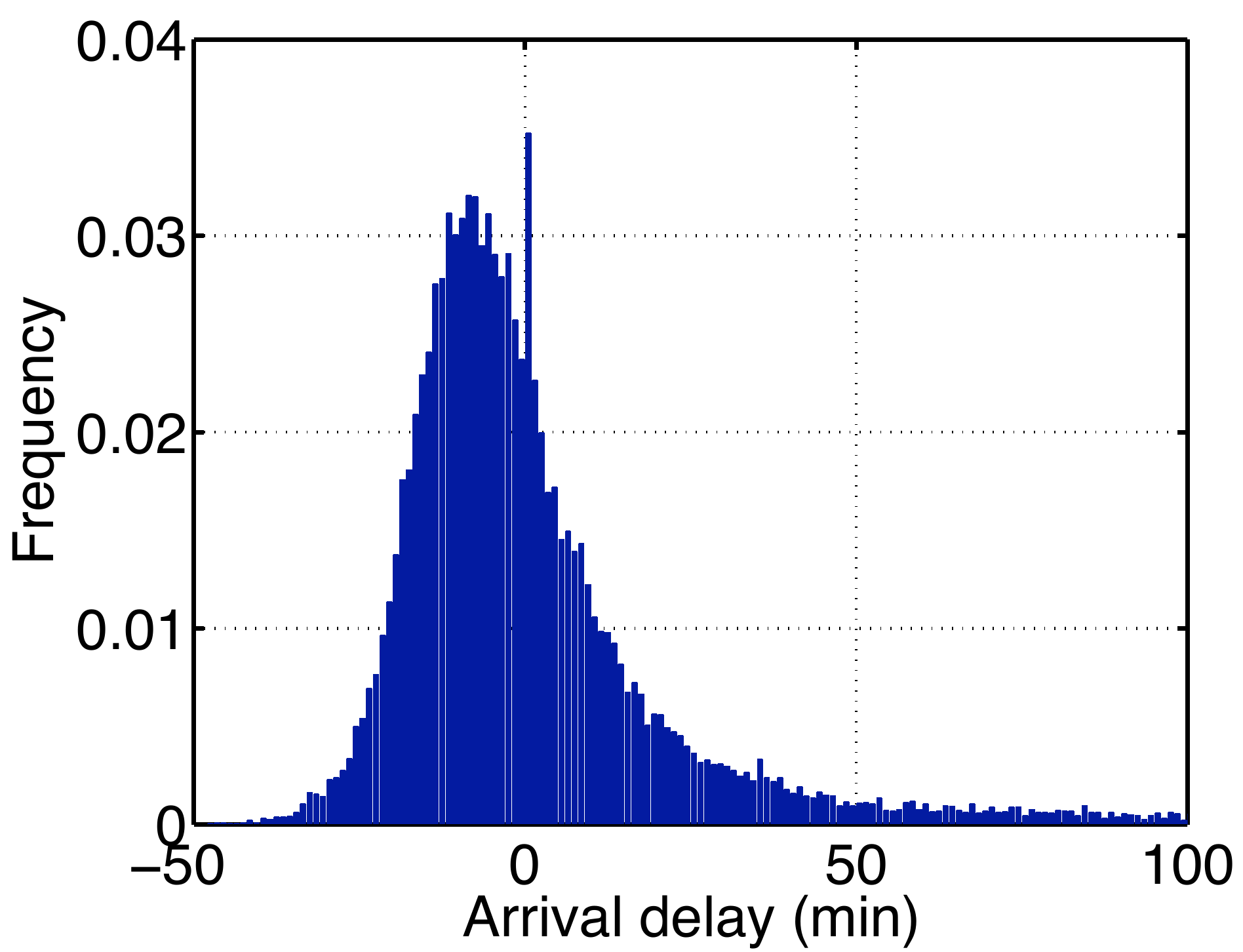}}
	\end{subfigmatrix}
	\caption{Delay distributions of DAL at ATL.}
	\label{f:atldelay}
\end{figure}

It is interesting that many departing flights departed earlier than scheduled as shown in Fig.~\ref{f:depdelay}. In order to see whether it happened only at DTW or not, gate delays of Delta Airlines (DAL) at Hartsfield-Jackson Atlanta International Airport (ATL) in May 2011 are shown in Fig.~\ref{f:atldelay}. It is shown that many departures at ATL also departed earlier than scheduled according to the airline on-time statistics provided by BTS \cite{bts}. 

\subsection{Delay Propagation Model}
In addition to the analysis of gate delays, the analysis of delay propagation in the turn around process is important to understand gate operations. An aircraft that arrives late is most likely to depart late, too. However, more precise knowledge of the interrelationship between arrival delay and departure delay is necessary. Shumsky studied the turn around process and suggested a turn model \cite{shumsky1995}. He assumed that a gate departure delay depends on ``available turn time.'' He defined the available turn time as the time remaining after an aircraft arrives at a gate until its scheduled departure time. So, the available turn time is $sch_d - act_a$, where $sch_d$ denotes the scheduled departure time and $act_a$ denotes the actual arrival time. The turn model proposed by Shumsky is given in Eq.~(\ref{e:turnmodel}). The quantity $dly_d$ is the departure delay, $C$ is a fixed amount of departure delay applied to every aircraft, $b$ is an additional departure delay ratio due to insufficient available turn time, $m$ is a minimum turn time, and $e$ is residual. According to Shumsky's model, there is a minimally required turn time, and an arrival delay is propagated to the corresponding departure delay when the available turn time is shorter than the minimum turn time ($m$). 
\begin{equation}
	\label{e:turnmodel}
	dly_d = C + b*\max(0,m-(sch_d - act_a)) + e.
\end{equation}

\begin{figure}[htb]
	\centering
	\includegraphics[width=0.7\textwidth]{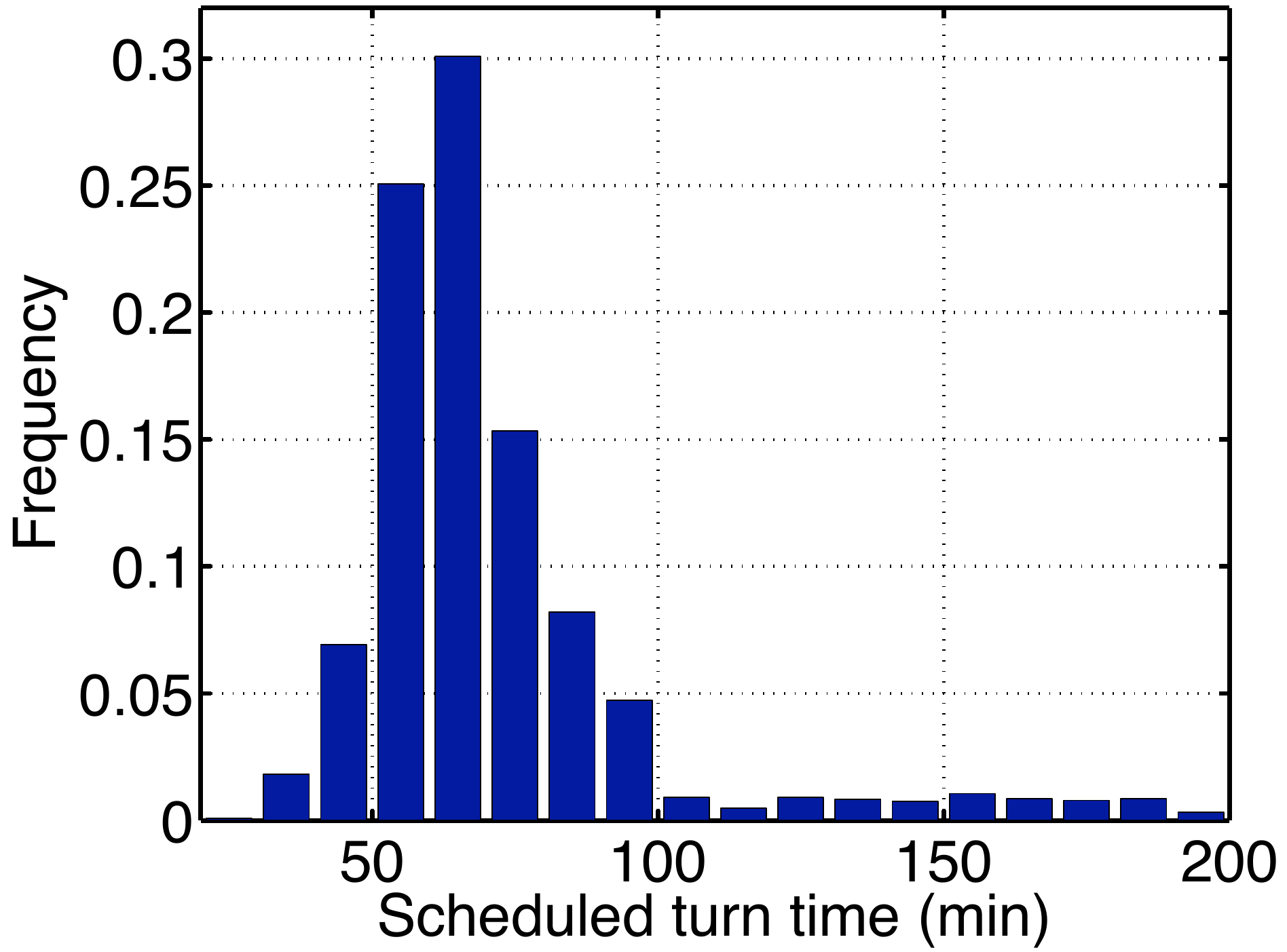}
	\caption{Distribution of scheduled turn times of NWA at DTW in March 2006.}
	\label{f:turntime}
\end{figure}

In order to analyze delay propagation, each arrival should be paired with the following departure that uses the same aircraft. The following departure is called the tail of the arrival. Because published data are separated into arrival data and departure data, tail numbers are used to identify aircraft. Applying this procedure to the NWA data at DTW in March 2006, 7545 arrival-departure pairs are identified, and 1072 pairs of them with a turn time shorter than 20 minutes or longer than 200 minutes are filtered out. It is considered that a turn time shorter than 20 minutes is not in normal operational conditions and a turn time longer than 200 minutes is not meaningful for the analysis of delay propagation because a departure delay after a large turn time (i.e., longer than 200 minutes) can be thought to be independent from the previous arrival delay. Then, 8 arrival-departure pairs are filtered out because their actual turn times are shorter than 20 minutes, which seems extraordinary. The distribution of scheduled turn times of 6465 arrival-departure pairs is shown in Fig.~\ref{f:turntime}. Most arrival-departure pairs are scheduled with turn times shorter than 100 minutes, and the majority of the turn times range from 50 minutes to 70 minutes.

\begin{figure}[htb]
	\centering
	\includegraphics[width=0.7\textwidth]{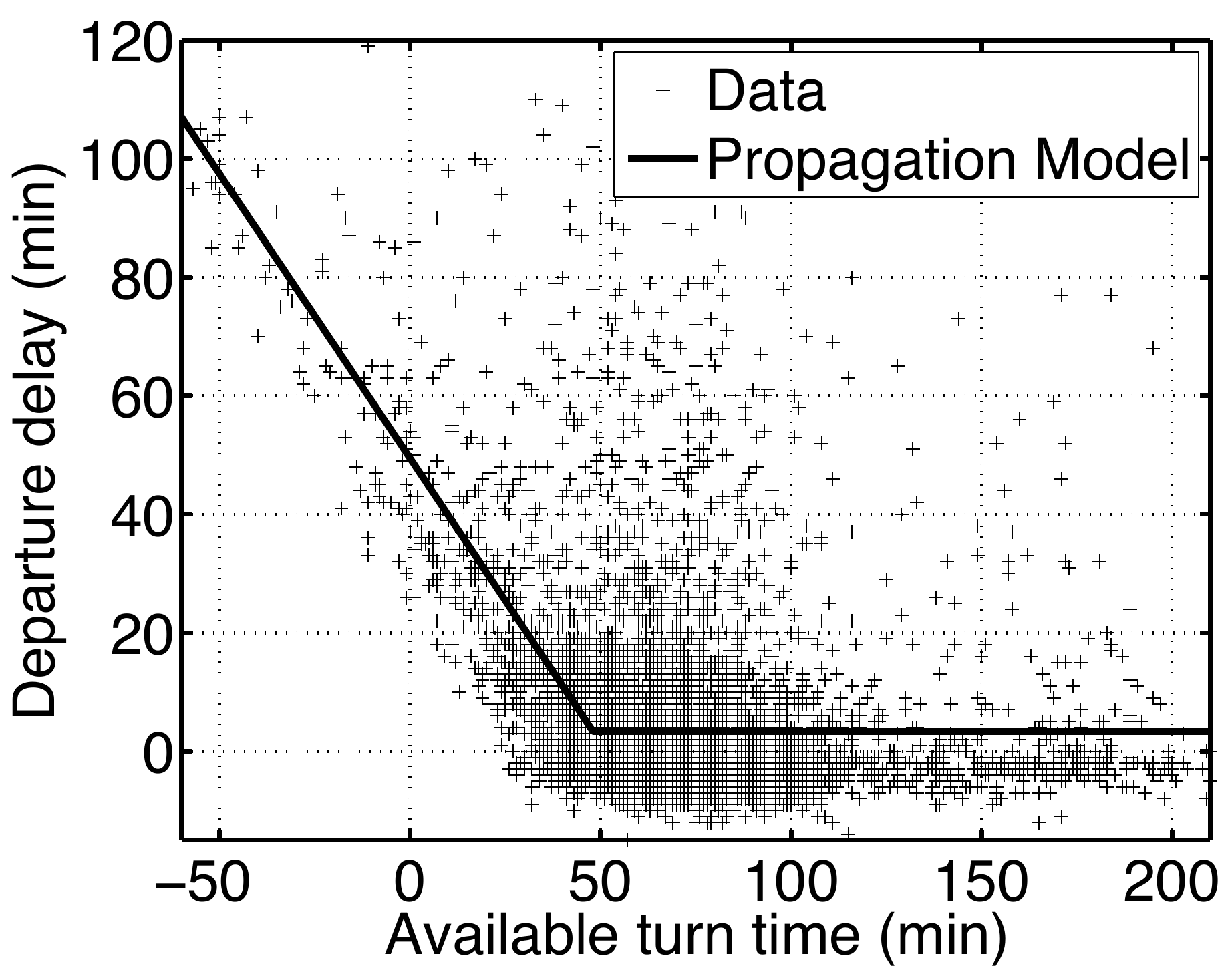}
	\caption{Delay propagation model for NWA at DTW in March 2006.}
	\label{f:propagation}
\end{figure}

Shumsky's model is applied to the data and shown in Fig.~\ref{f:propagation}. Parameters of the delay propagation model are given in Table~\ref{t:turnmodel}. It is shown that the minimum required turn time of the data is 48 minutes and every minute of an arrival delay above the minimum is propagated to 0.96 minute of a departure delay.

\begin{table}[htb]
	\begin{center}
	\caption{Parameters of delay propagation model for NWA at DTW in March 2006}
	\label{t:turnmodel}
	\begin{tabularx}{0.5\textwidth}{XXX}
		\hline
		$M$ & $C$ & $b$ \\ \hline
		48 & 3.379 & 0.96 \\
		\hline
	\end{tabularx}
	\end{center}
\end{table}

\section{Robust Gate Assignment}
\subsection{Calculation of Gate Conflict Duration}
\begin{figure}[htb]
	\centering
	\includegraphics[width=0.7\textwidth]{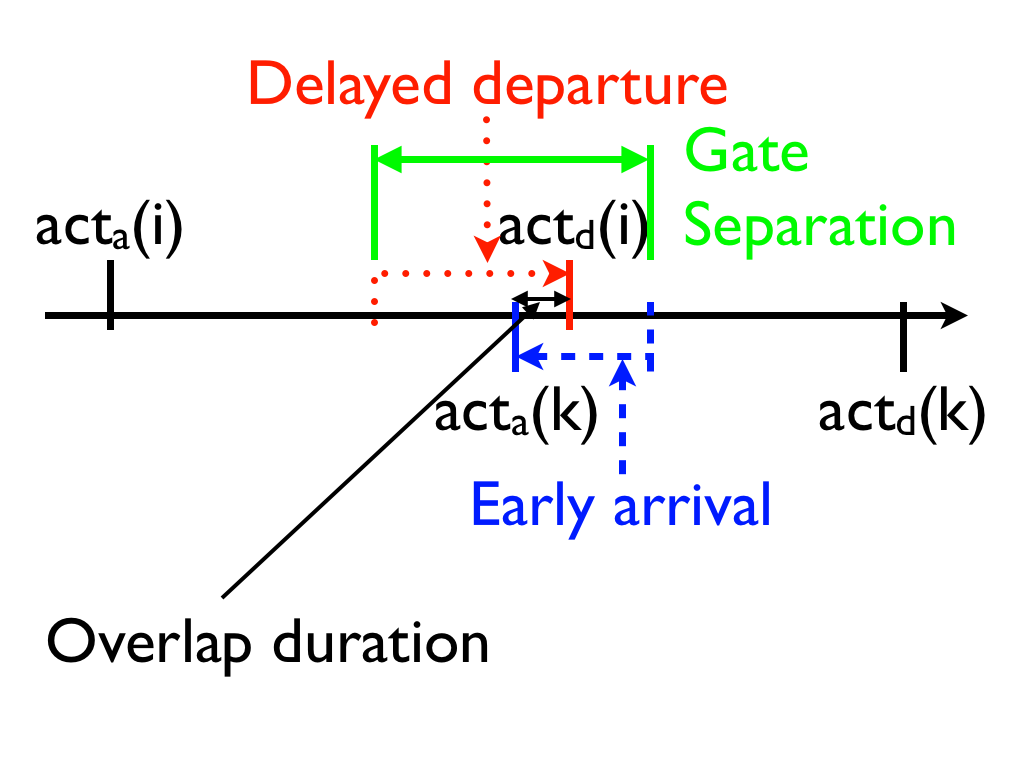}
	\caption{Typical gate conflict where two aircraft need the same gate at the same time: Flight $i$ is scheduled to depart before flight $k$ arrives at the gate, but the flight $k$ arrives before the flight $i$ pushes back.}
	\label{f:gateconflict}
\end{figure}

The objective of robust gate assignment is to maximize the robustness of gate assignments. ``Robust'' means that the gate assignment is resistant to uncertain delays. Indeed, severe delays perturb gate operations by forcing arriving aircraft to wait for gates, or ramp controllers to reassign gates. The disturbances can be reduced if the gate assignment is robust against uncertain delays. Therefore, the objective is to minimize the duration of gate conflicts, equivalently. If a gate is still occupied by an aircraft when another aircraft requests the gate, the latter should wait until the assigned gate or another gate is available (gate conflict). Fig.~\ref{f:gateconflict} illustrates a gate conflict, where $act_a(i)$ and $act_d(i)$ denote the actual arrival time and the actual departure time of flight $i$. The gate separation is the time gap between the scheduled departure time of flight $i$ ($sch_d(i)$, red dashed line) and the scheduled arrival time of flight $k$ ($sch_a(k)$, blue dashed line). In Fig.~\ref{f:gateconflict}, flight $i$ is scheduled to leave the gate before flight $k$ arrives, but the departure time of flight $i$ is delayed and flight $k$ arrives earlier than scheduled. So, when flight $k$ arrives, the gate is not released yet and flight $k$ has to wait for $act_d(i)-act_a(k)$.

Because the actual arrival time ($act_a$) and the actual departure time ($act_d$) are unknown when gates are assigned, the duration of a gate conflict is estimated based on the probability distributions of arrival delay and departure delay. As shown in the previous section, arrival delay ($dly_a$) and departure delay ($dly_d$) are modeled using a Log-normal distribution. The actual departure time of flight $i$ and the actual arrival time of flight $k$ are given in Eqs.~(\ref{e:actdep})-(\ref{e:actarr}), where $\mu$ and $\sigma$ are parameters of the Log-normal distribution and $c$ is the shift parameter.
\begin{align}
	\label{e:actdep}
	&act_d(i) = sch_d(i) + dly_d(i) = sch_d(i) + \{ \mathbf{Log-}\mathcal{N}(\mu_d, \sigma_d) + c_d \}, \\
	\label{e:actarr}
	&act_a(k) = sch_a(k) + dly_a(k) = sch_a(k) + \{ \mathbf{Log-}\mathcal{N}(\mu_a, \sigma_a) + c_a \}.
\end{align}

In Fig.~\ref{f:gateconflict}, the duration of gate conflict is $act_d(i) - act_a(k)$. Because the actual times (i.e., $act_d(i)$ and $act_a(k)$) are random variables, the expected duration of gate conflict is calculated below.
\begin{align}
	\label{e:disturb}
	&\text{Expected duration of gate conflict} = \mbox{E}[act_d(i)-act_a(k), act_d(i)>act_a(k)] \\
	\label{e:disturb1}
	&= \mbox{E}[sch_d(i) + dly_d(i) - sch_a(k) - dly_a(k), dly_d(i) - dly_a(k) > sch_a(k) - sch_d(i)] \\
	\label{e:disturb2}
	&= \mbox{E}[\mathbf{Log-}\mathcal{N}(\mu_d, \sigma_d) - \mathbf{Log-}\mathcal{N}(\mu_a, \sigma_a) - z, \mathbf{Log-}\mathcal{N}(\mu_d, \sigma_d) - \mathbf{Log-}\mathcal{N}(\mu_a, \sigma_a) > z ] \\
	\label{e:disturb3}
	&= \int_{0}^{\infty} \int_{y+z}^{\infty} (x-y-z) f_{dep}(x; \mu_d, \sigma_d, 0) f_{arr}(y; \mu_a, \sigma_a, 0) dx dy, \\
	& \text{where } z = sch_a(k) - sch_d(i) - c_d + c_a. \nonumber
\end{align}

\begin{figure}[htb]
	\centering
	\includegraphics[width=0.7\textwidth]{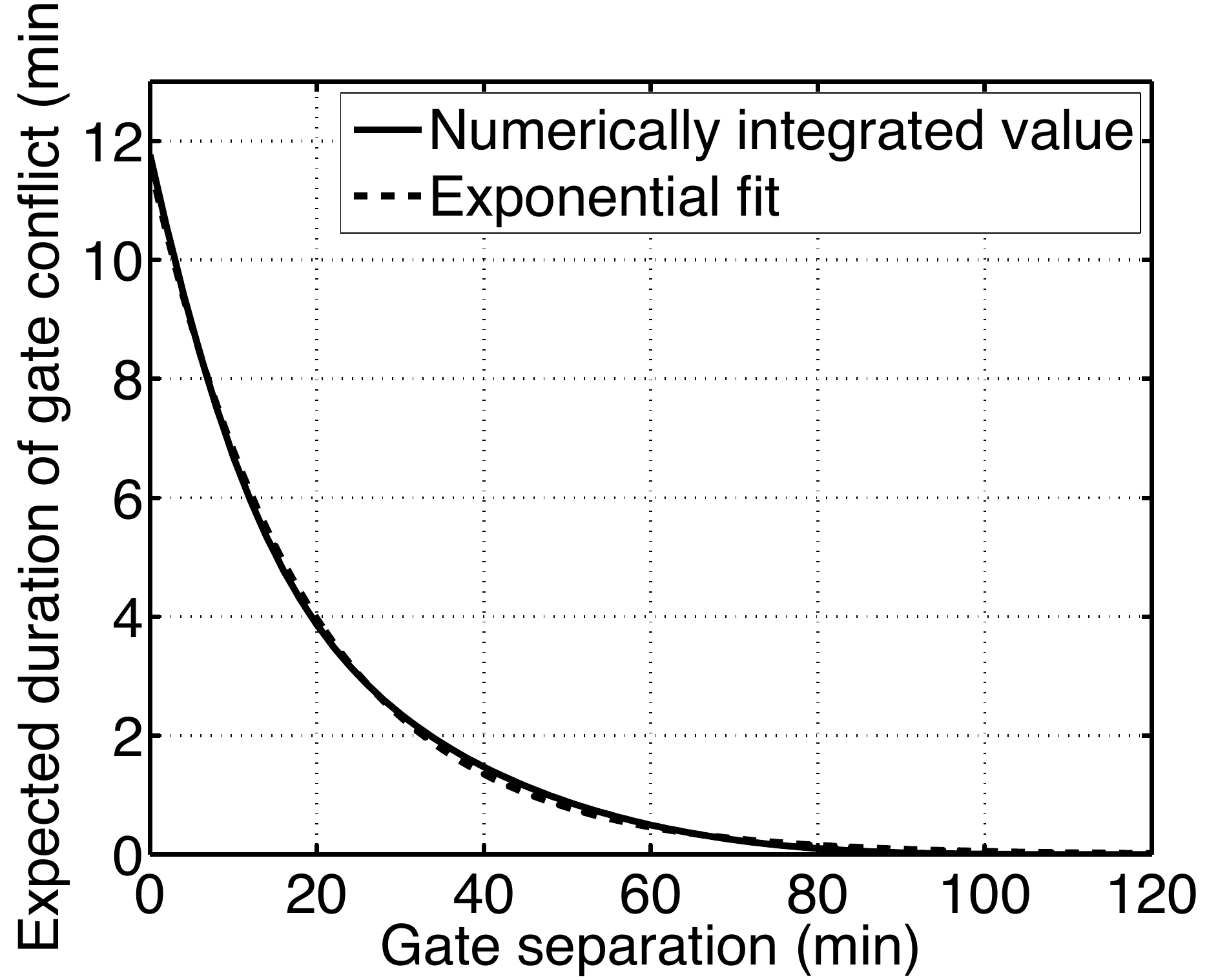}
	\caption{Expected duration of gate conflict as a function of planned gate separation between consecutive occupancies, together with the exponential fit $11.6 \times 0.95^{\text{gate separation}}$: The expected duration of gate conflict decays exponentially as gate separation increases.}
	\label{f:duration}
\end{figure}

Using Eqs.~(\ref{e:actdep})-(\ref{e:actarr}), Eq.~(\ref{e:disturb}) is formulated to Eqs.~(\ref{e:disturb1}) and (\ref{e:disturb2}). It is assumed that the departure delay of flight $i$ and the arrival delay of flight $k$ are independent. So, Eq.~(\ref{e:disturb2}) is calculated by a double integral using the PDF of departure delays ($f_{dep}$) and that of arrival delays ($f_{arr}$). The PDF is given in Eq.~(\ref{e:lognpdf}). Eq.~(\ref{e:disturb3}) depends only on $z$, or equivalently $sch_a(k) - sch_d(i)$, which is the gate separation between flight $i$ and flight $k$. Because there is no closed form for Eq.~(\ref{e:disturb3}), the double integral is calculated numerically, and Fig.~\ref{f:duration} shows the expected duration of gate conflicts according to gate separations. Note that the expected duration of gate conflicts is only about 12 minutes when gate separation is zero. The duration is surprisingly small because early departures and late arrivals occur frequently as shown in Fig.~\ref{f:delay}.

Because the curve decreases exponentially and the calculation of Eq.~(\ref{e:disturb3}) is time-consuming, the curve is fitted to an exponential function given in Eq.~(\ref{e:disturb4}), where $\text{sep}(i,k)$ denotes the gate separation between flight $i$ and flight $k$ as given in Eq.~(\ref{e:separation}). The parameter $a$ is the $y$-intercept of the exponential function and $b$ is the exponential base. For the NWA data at DTW in March 2006, $a$ and $b$ are 11.63 and 0.9476, respectively. As shown in Fig.~\ref{f:duration}, the exponential function models the expected duration of gate conflicts well.
\begin{align}
	\label{e:disturb4}
	&\text{Expected duration of gate conflict} \sim a \times b^{\text{sep}(i, k)}. \\
	\label{e:separation}
	&\text{sep}(i, k) = \left\{ 
		\begin{array}{rl}
		sch_a(k) - sch_d(i) & \text{if flight } i \text{ is followed by flight } k \\
		sch_a(i) - sch_d(k) & \text{otherwise.}
		\end{array} \right.
\end{align}

\subsection{Problem Formulation}
\begin{align}
 \label{e:rga}
 &\text{Minimize} \sum_{i \in \mathcal{F}} \sum_{k \in \mathcal{F}, k > i} a\times b^{\text{sep}(i,k)} \sum_{j \in \mathcal{G}} \ x_{ij} \ x_{kj} \\
 &\nonumber \text{subject to} \\
 \label{e:gateconst}
 &\sum_{j \in \mathcal{G}} x_{ij} = 1, \ \forall i \in \mathcal{F} \\
 \label{e:feasconst}
 &(sch_d(i) - sch_a(k) + t^{\mbox{buff}}) (sch_d(k) - sch_a(i) + t^{\mbox{buff}}) \leq M (2 - x_{ij} - x_{kj}) \nonumber \\
 &, \ i \neq k, \ \forall i, k \in \mathcal{F}, \ \forall j \in \mathcal{G} \\
 \label{e:decision}
 &x_{ij} \in \{0,1\}, \ \forall i \in \mathcal{F}, \ \forall j \in \mathcal{G}, \\
 &\nonumber \text{where } x_{ij} = \left\{ 
  \begin{array}{rl}
   1 & \text{if flight } i \text{ is assigned to gate } j \\
   0 & \text{otherwise.}
   \end{array} \right.
\end{align}

A quadratic integer formulation of robust gate assignment problem is given in Eqs.~(\ref{e:rga})-(\ref{e:decision}). The decision variable $x_{ij}$ indicates whether flight $i$ is assigned to gate $j$. The sets $\mathcal{F}$ and $\mathcal{G}$ denote the sets of flights and gates, respectively. As shown in Eq.~(\ref{e:disturb4}), the expected duration of gate conflicts depends on the gate separation. Hence, only if flight $i$ and flight $k$ use the same gate, the expected duration of gate conflict between them contributes to the objective function. In other words, there is no gate conflict between flight $i$ and flight $k$ when their gate assignments are different (i.e., $x_{ij} \ x_{kj} = 0$). 

\begin{figure}[htb]
\begin{subfigmatrix}{2}
 \subfigure[Feasible gate schedule with sufficient gate separation.]{\label{f:feasible}\includegraphics{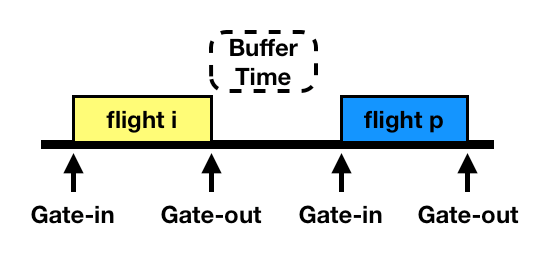}}
 \subfigure[Infeasible gate schedule with insufficient gate separation.]{\label{f:infeasible}\includegraphics{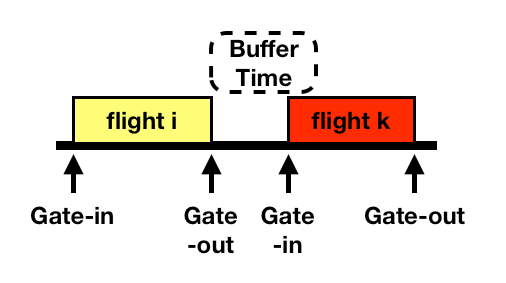}}
\end{subfigmatrix}
 \caption{Feasible and infeasible gate schedules.}
 \label{f:feas}
\end{figure}

Two constraints are given in Eq.~(\ref{e:gateconst}) and Eq.~(\ref{e:feasconst}). The first constraint makes sure that every flight is assigned to a single gate. The second constraint constrains two successive gate occupancies, so that they are separated by more than a certain amount of time, which is called buffer time ($t^{\mbox{buff}}$). Note that scheduled times ($sch_a$ and $sch_d$) are used for gate assignment. The buffer time is used to absorb disturbances of schedule, e.g. stochastic delays, and to ensure the feasibility of the schedule. Two series of gate schedules with different gate separations are illustrated in Fig.~\ref{f:feas}. The gate schedules in Fig.~\ref{f:feasible} have a gate separation longer than the buffer time. So, flight $i$ and flight $p$ can be assigned to the same gate. On the other hand, the gate separation in Fig.~\ref{f:infeasible} is shorter than the buffer time. Hence, flight $i$ and flight $k$ cannot be assigned to the same gate. Note that Eq.~(\ref{e:feasconst}) is binding if and only if flight $i$ and flight $k$ are assigned to gate $j$ (i.e., $x_{ij}=x_{kj}=1$) because $M$ is an arbitrarily large number. Our previous research shows that Tabu Search (TS) outperforms Branch and Bound (B\&B) and Genetic Algorithm (GA) in solution time and solution accuracy \cite{kim2012aga}. Therefore, the proposed problem is solved by TS.

\section{Results}
\subsection{Analysis of Robust Gate Assignment}
We propose an approach for robust gate assignment. In order to evaluate the proposed approach, we select NWA flight schedules on March 1st, 2006, at DTW for the baseline schedule (1x schedule). The baseline gate assignments are available on the internet \cite{flightstats}, and 226 arrival-departure pairs are found excluding flights that stay overnight at the airport. In addition to the baseline schedule, we generate three more flight schedules with increased traffic (1.1x, 1.2x, and 1.3x) in order to analyze the impact of the traffic volume on the robustness of gate assignment. The number of gates is 44 for all the flight schedules.  

The baseline schedule and other schedules are assigned to gates based on the proposed robust gate assignment and a greedy policy. The robust gate assignment minimizes the sum of expected gate conflict duration, and the greedy policy tries to pack flight schedules as much as possible. So, the greedy policy works against the robustness of gate assignment, and we will analyze the impact of the gate assigning policy on the robustness of gate assignment.

\begin{table}[htb]
 \begin{center}
  \caption{Comparison of gate assignments}
  \label{t:assign}
  \begin{tabularx}{\textwidth}{lXXX}
	\hline
	Assignment & Total expected gate conflict duration & Number of utilized gates & Minimum gate separation \\ \hline
	Baseline & N/A & 44 & 3 min\\
	TS (1x schedule) & 23.8 min & 44 & 42 min\\
	Greedy (1x schedule) & 296.5 min & 38 & 15 min\\ \hline
	TS (1.1x schedule) & 33.4 min & 44 & 31 min\\
	Greedy (1.1x schedule) & 535.8 min & 37 & 15 min\\ \hline
	TS (1.2x schedule) & 51.9 min & 44 & 25 min\\
	Greedy (1.2x schedule) & 549.6 min & 41 & 15 min\\ \hline
	TS (1.3x schedule) & 193.7 min & 44 & 15 min\\
	Greedy (1.3x schedule) & 589.0 min & 44 & 15 min \\
	\hline
  \end{tabularx}
 \end{center}
\end{table}

The resulting gate assignments including the baseline gate assignment are given in Table~\ref{t:assign}. Because the greedy policy assigns gates as compactly as possible, the corresponding assignments result in the highest duration of gate conflicts. As shown in Fig.~\ref{f:duration}, the shorter the gate separation, the higher the duration of gate conflict. So, the minimum gate separation is an easy parameter for the worst case (most sensitive flight to a delay) of a gate assignment.

\begin{figure}[htb]
\begin{subfigmatrix}{2}
 \subfigure[Total duration of gate conflict.]{\label{f:simul1}\includegraphics{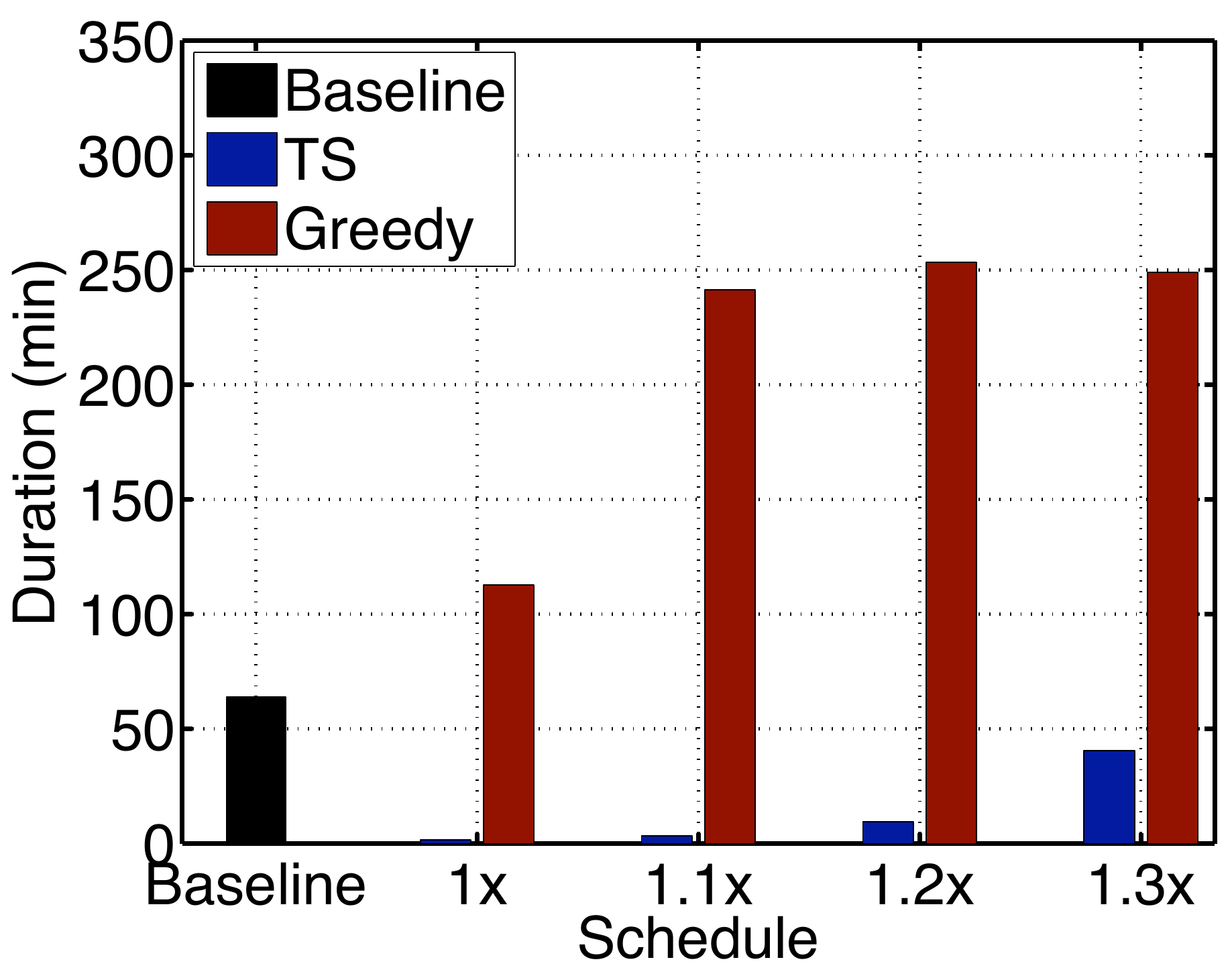}}
 \subfigure[Number of gate conflicts.]{\label{f:simul2}\includegraphics{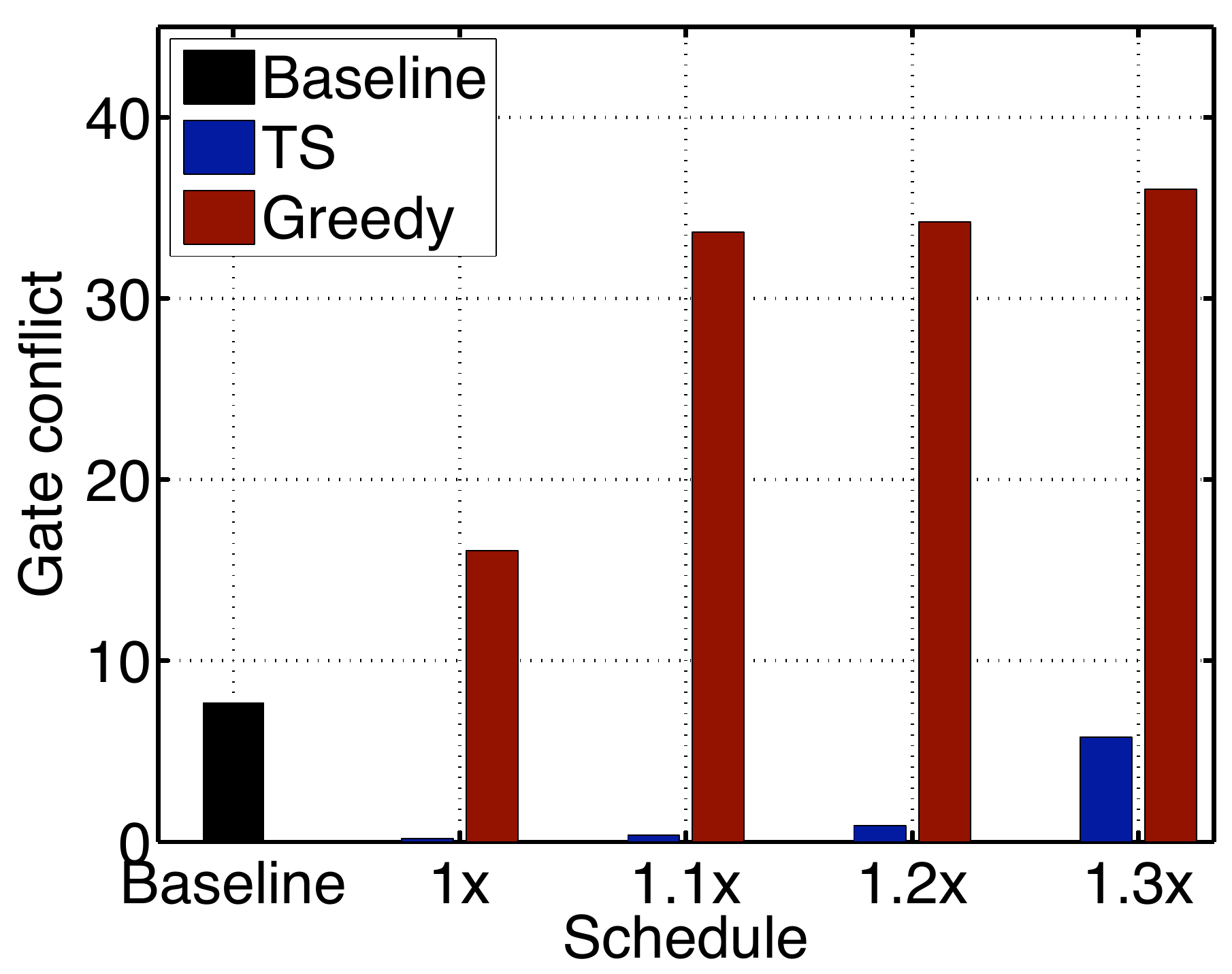}}
\end{subfigmatrix}
 \caption{Simulation results.}
 \label{f:simulation}
\end{figure}

\begin{table}[htb]
 \begin{center}
  \caption{Comparison of the baseline assignment and the robust assignment}
  \label{t:comparison}
  \begin{tabularx}{\textwidth}{lXX}
	\hline
	Assignment & Duration of gate conflict & Number of gate conflicts \\ \hline
	Baseline & 62.1 min & 7.67\\
	TS (1x schedule) & 2.27 min & 0.25 \\
	\hline
  \end{tabularx}
 \end{center}
\end{table}

Actual arrival and departure times are simulated in order to evaluate gate assignments in Table~\ref{t:assign}. Actual arrival times are calculated by Eq.~(\ref{e:actarr}) and departure delays are generated by the delay propagation model given in Eq.~(\ref{e:turnmodel}). 100 simulation runs are executed for each assignment. If a gate conflict occurs, the arrival time of the following flight is delayed until the preceding flight departs. Fig.~\ref{f:simul1} shows the total duration of gate conflicts, and Fig.~\ref{f:simul2} shows the number of gate conflicts. It is shown that the robust gate assignment (TS) is more robust than the baseline gate assignment even when the traffic volume is increased by 30\%. On the other hand, dense gate assignment by Greedy policy is harmful to the robustness of gate assignments. Table~\ref{t:comparison} compares the robust gate assignment to the baseline gate assignment with the current traffic volume (1x schedule). The robust gate assignment reduces the duration of gate conflicts by 96.3\% and the number of gate conflicts by 96.7\% compared to the baseline assignment. 

\begin{figure}[htb]
\begin{subfigmatrix}{2}
 \subfigure[Gate conflict duration per aircraft.]{\label{f:simul3}\includegraphics{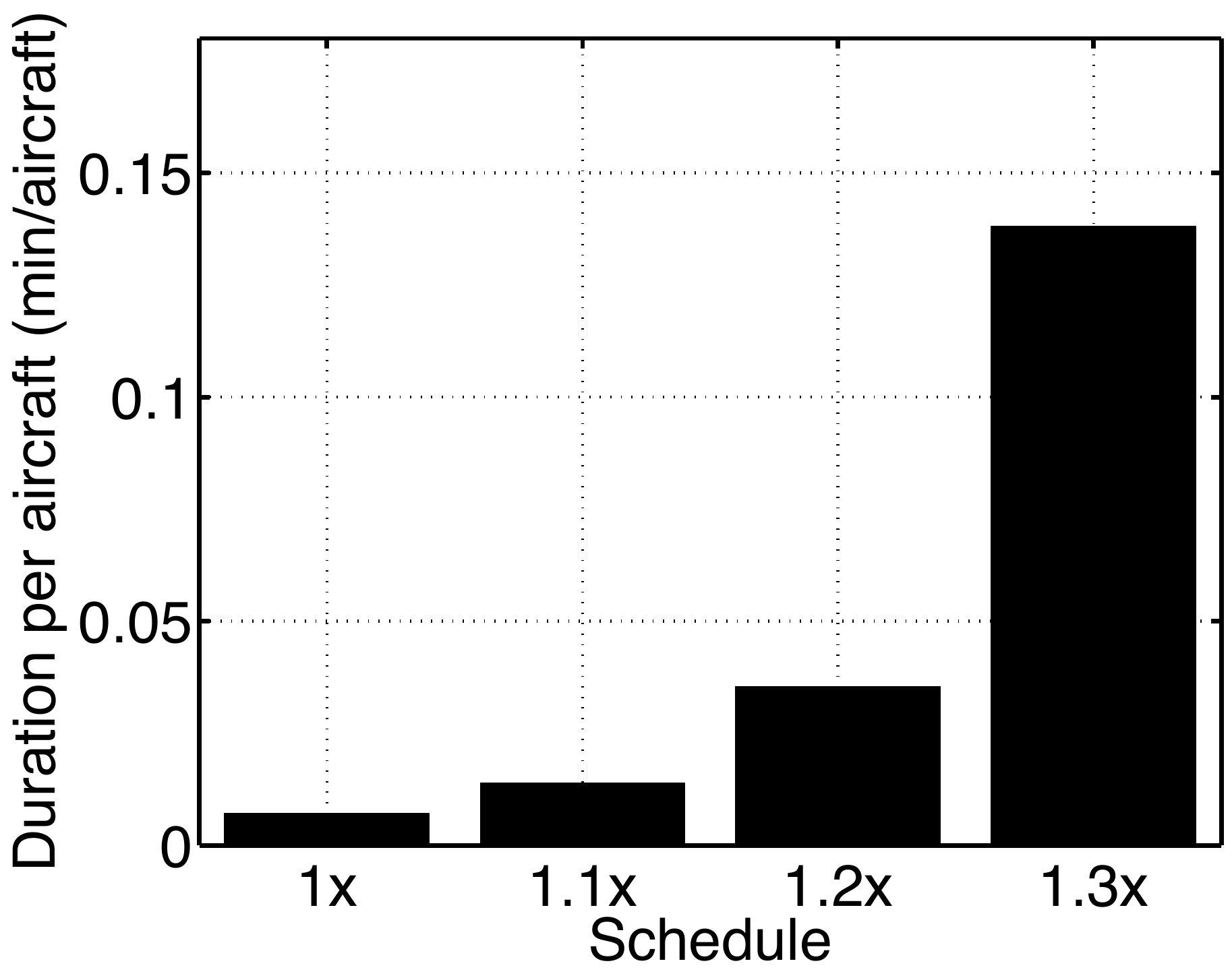}}
 \subfigure[Number of gate conflicts per aircraft.]{\label{f:simul4}\includegraphics{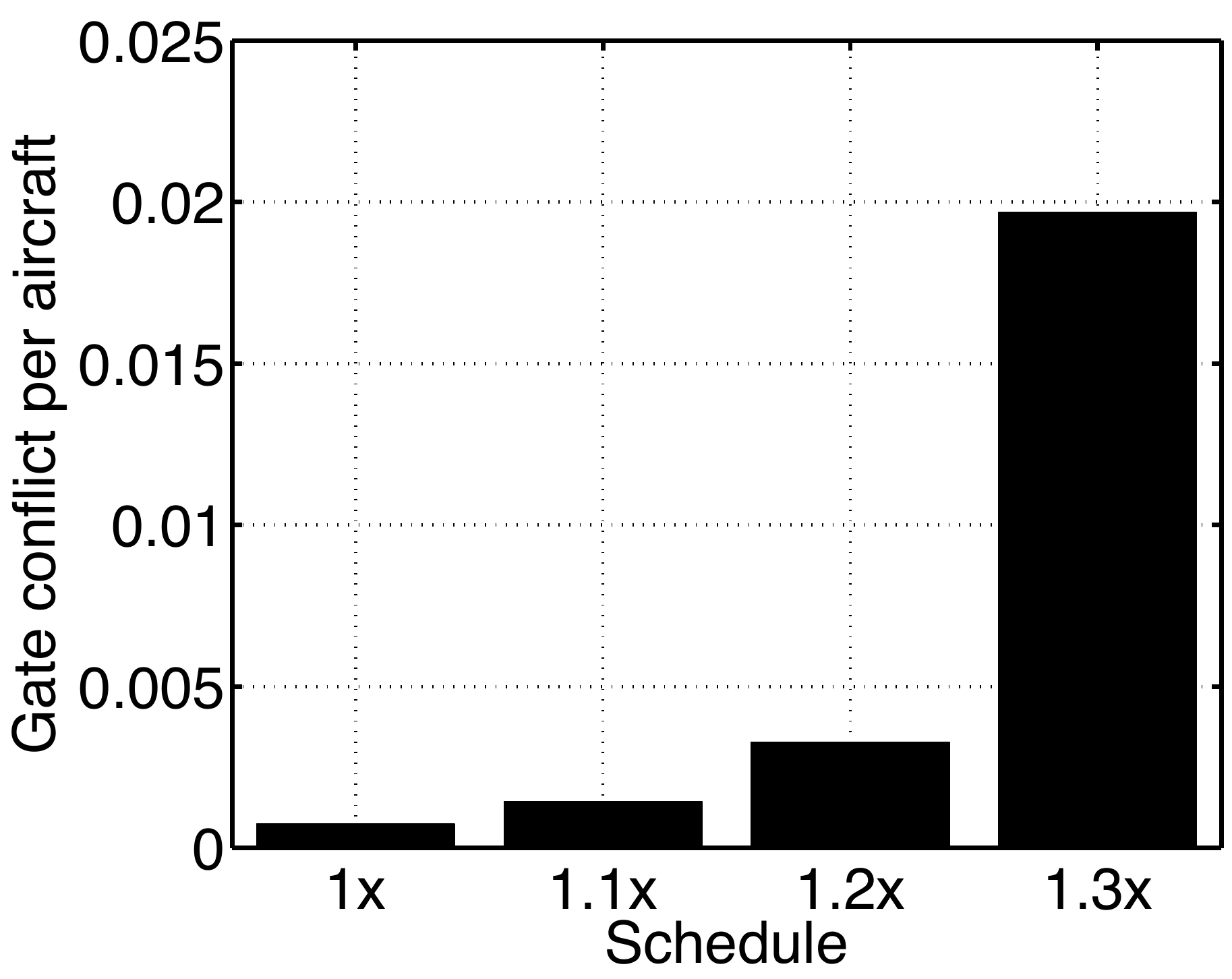}}
\end{subfigmatrix}
 \caption{Normalized results (assigned by TS).}
 \label{f:simulation2}
\end{figure}

\begin{figure}[htb]
 \centering
 \includegraphics[width=0.7\textwidth]{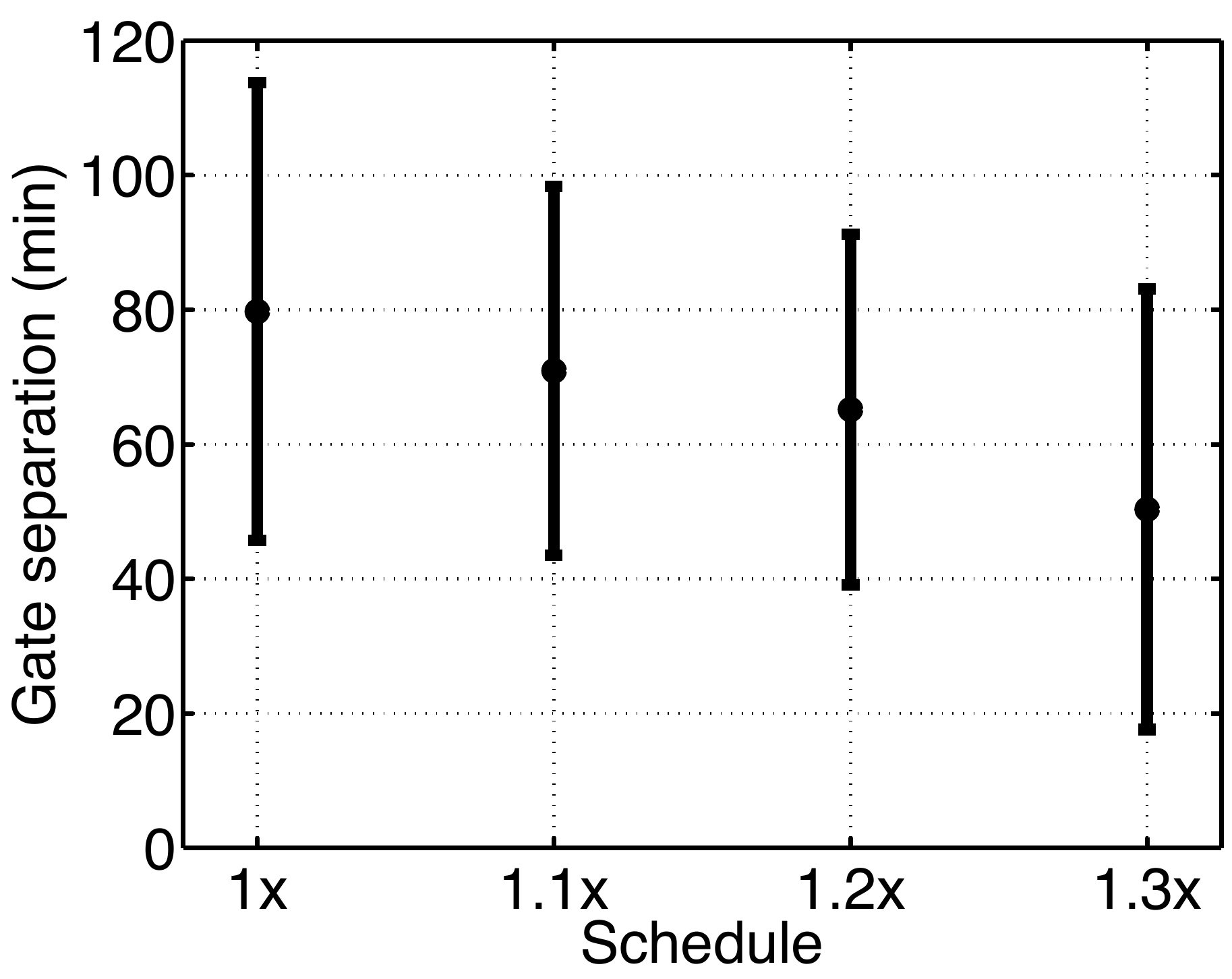}
 \caption{Average and standard deviation of gate separations (assigned by TS): Circles indicate the mean gate separation and bars shows one standard deviation.}
 \label{f:sep}
\end{figure}

To analyze the influence of increased traffic on the robustness of gate assignment, the total gate conflict duration and the number of gate conflicts are normalized by the number of aircraft and shown in Fig.~\ref{f:simulation2}. The duration and the number increase highly as traffic increases. It is explained by the relationship between gate separation and gate conflict duration shown in Fig.~\ref{f:duration}: Fig.~\ref{f:sep} compares the mean and the standard deviation of gate separations of each gate assignment. Because the number of gates is constant while traffic increases from 1x to 1.3x, gate assignment of 1.3x traffic is denser than that of 1x traffic. More specifically, the mean gate separation decreases by 36.7\% when traffic increases by 30\% so the duration and the number of gate conflict increase. 

\subsection{Impact of Robust Gate Assignment on Passenger Transit Time}
\begin{figure}[htb]
 \centering
 \includegraphics[width=0.5\textwidth]{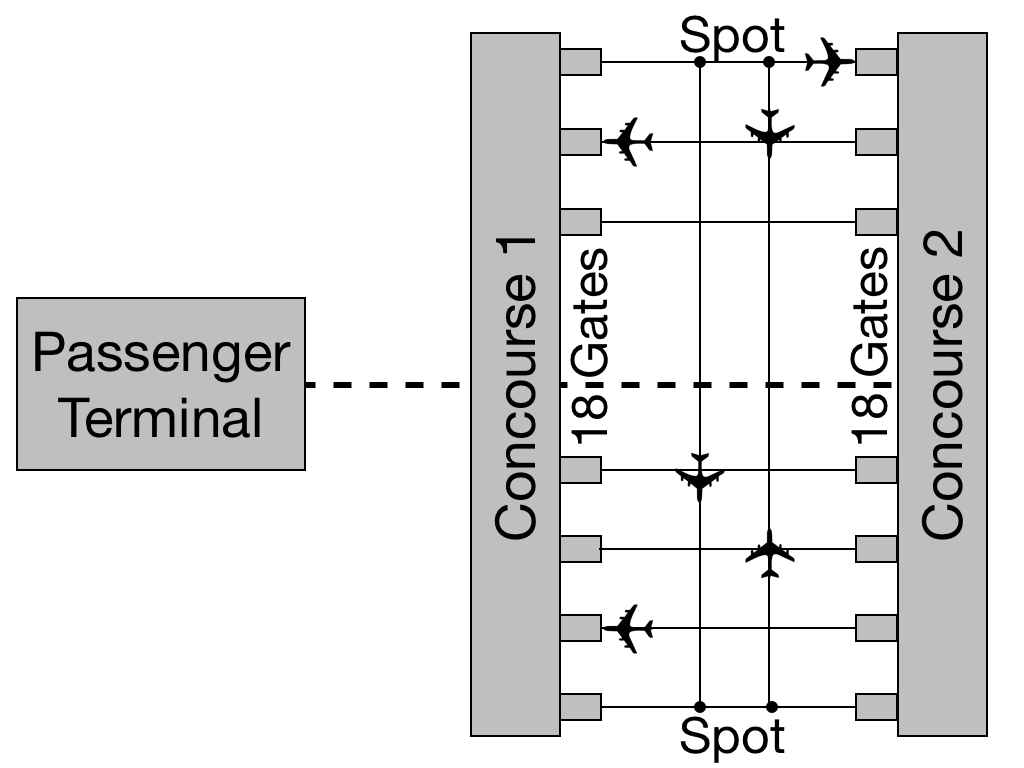}
 \caption{Example 1: Parallel ramp configuration}
 \label{f:ramp1}
\end{figure}

The robust gate assignment tends to distribute flight schedules over gates in order to disperse gate separations. As a consequence, connection flights are also likely to be scattered over gates and passenger transit time increases. Because passenger connection data for the previous example of NWA is not available, we use fictitious examples in our previous study \cite{kim2012aga}. Two examples of ramp configuration are presented, and 200 flights and passenger data are randomly generated. Fig.~\ref{f:ramp1} shows the first example ramp with parallel concourses. This example is a simplified Hartsfield-Jackson Atlanta Airport, and there are two parallel concourses that have 18 gates. Two concourses and passenger terminal are connected by an underground people mover.

\begin{align}
 \label{e:rga2}
 &\text{Obj}_\text{robust} = \sum_{i \in \mathcal{F}} \sum_{k \in \mathcal{F}, k > i} a\times b^{\text{sep}(i,k)} \times n^{in} \sum_{j \in \mathcal{G}} \ x_{ij} \ x_{kj}, \\
 \label{e:rga3}
 &\text{Obj}_\text{transit} = \sum_{i \in \mathcal{F}} \sum_{j \in \mathcal{G}} (n^{\mbox{o}}_i \times \frac{d^{\mbox{s}}_j}{v^{\mbox{m}}} + n^{\mbox{d}}_i \times \frac{d^{\mbox{b}}_j}{v^{\mbox{m}}}) \ x_{ij} + \sum_{i \in \mathcal{F}} \sum_{j \in \mathcal{G}} \sum_{k \in \mathcal{F}, k > i} \sum_{l \in \mathcal{G}} n_{ik} \times \frac{d_{jl}}{v^{\mbox{m}}} \ x_{ij} \ x_{kl}.
\end{align}

In order to compare the robustness of gate assignment with passenger transit time, the expected duration of gate conflict in Eq.~(\ref{e:rga}) is weighted by the number of arrival passengers ($n^{\mbox{in}}$) because only arrivals are delayed due to a gate conflict. The modified objective function of the robust gate assignment is given in Eq.~(\ref{e:rga2}), and the objective function of passenger transit time is given in Eq.~(\ref{e:rga3}), which is borrowed from our previous study \cite{kim2012aga}. The quantity $n^{\mbox{o}}_i$ is the number of origin passengers of flight $i$, $n^{\mbox{d}}_i$ is the number of destination passengers of flight $i$, and $n_{ik}$ denote the number of transfer passengers between flight $i$ and flight $k$. The distance from a security checkpoint to a gate $j$ is $d^{\mbox{s}}_j$, the distance from a gate $j$ to a baggage claim is $d^{\mbox{b}}_j$, and the distance between two gates is $d_{jl}$. The quantity $v^{\mbox{m}}$ denotes the average moving speed of passengers, which varies with the configuration of passenger terminal: $v^{\mbox{m}}$ is higher where passengers can move faster by taking moving walkways, underground people mover, etc. In order to analyze the trade-off, the objective function is given as a linear combination of Eq.~(\ref{e:rga2}) and Eq.~(\ref{e:rga3}) as shown in Eq.~(\ref{e:tradeobj2}).
\begin{equation}
	\label{e:tradeobj2}
	\text{Obj} = (1-\alpha) \times \text{Obj}_\text{transit} + \alpha \times \text{Obj}_\text{robust}.
\end{equation}
A trade-off factor $\alpha$ is introduced. When $\alpha$ is 0, the resulting optimization problem minimizes only the passenger transit time. When $\alpha$ is 1, minimizing the duration of gate conflicts is the only objective of the optimization problem. 

\begin{figure}[htb]
\begin{subfigmatrix}{3}
 \subfigure[$\alpha$ versus total transit time.]{\label{f:transit1}\includegraphics{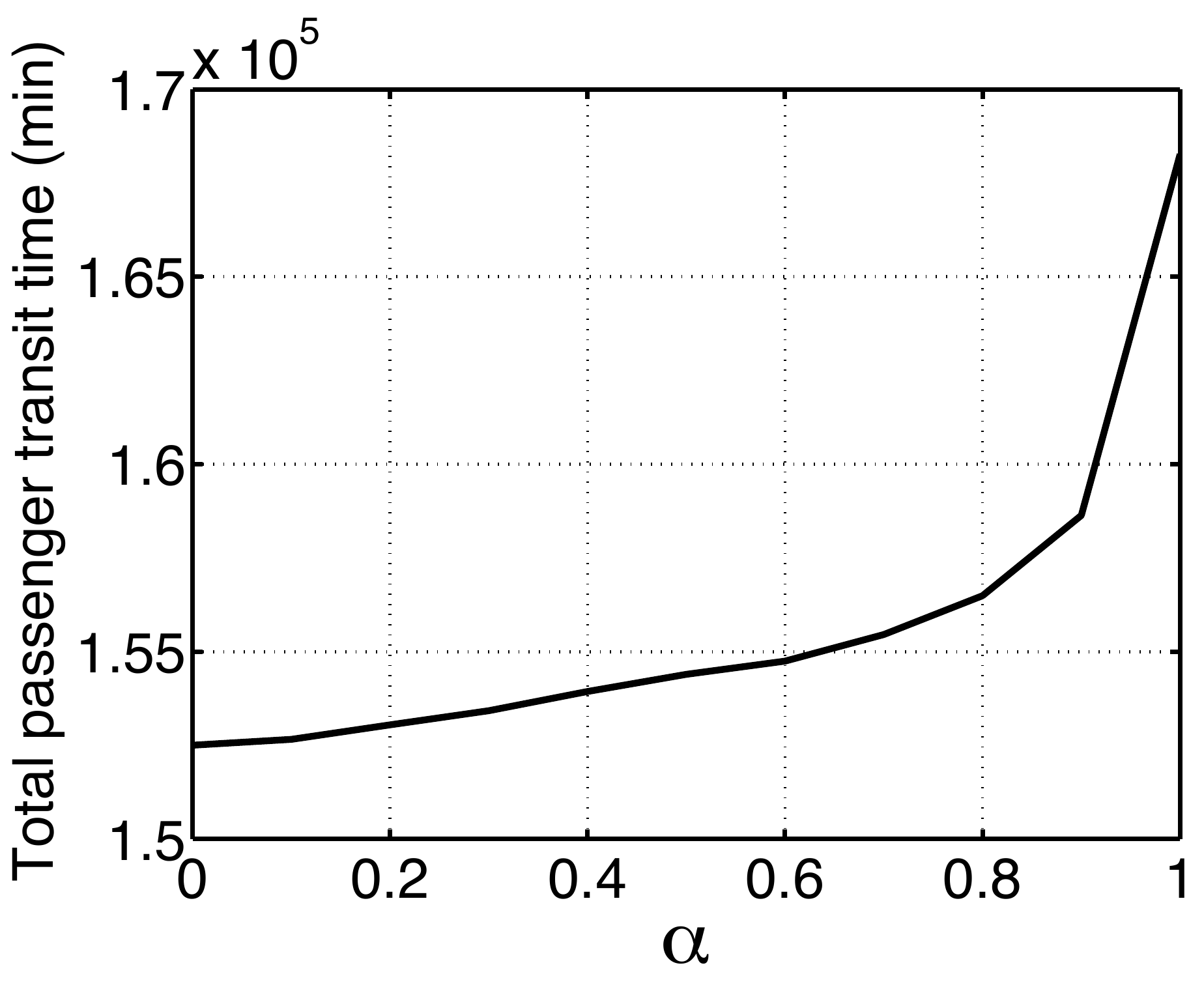}}
 \subfigure[$\alpha$ versus total weighted gate conflict duration.]{\label{f:robust1}\includegraphics{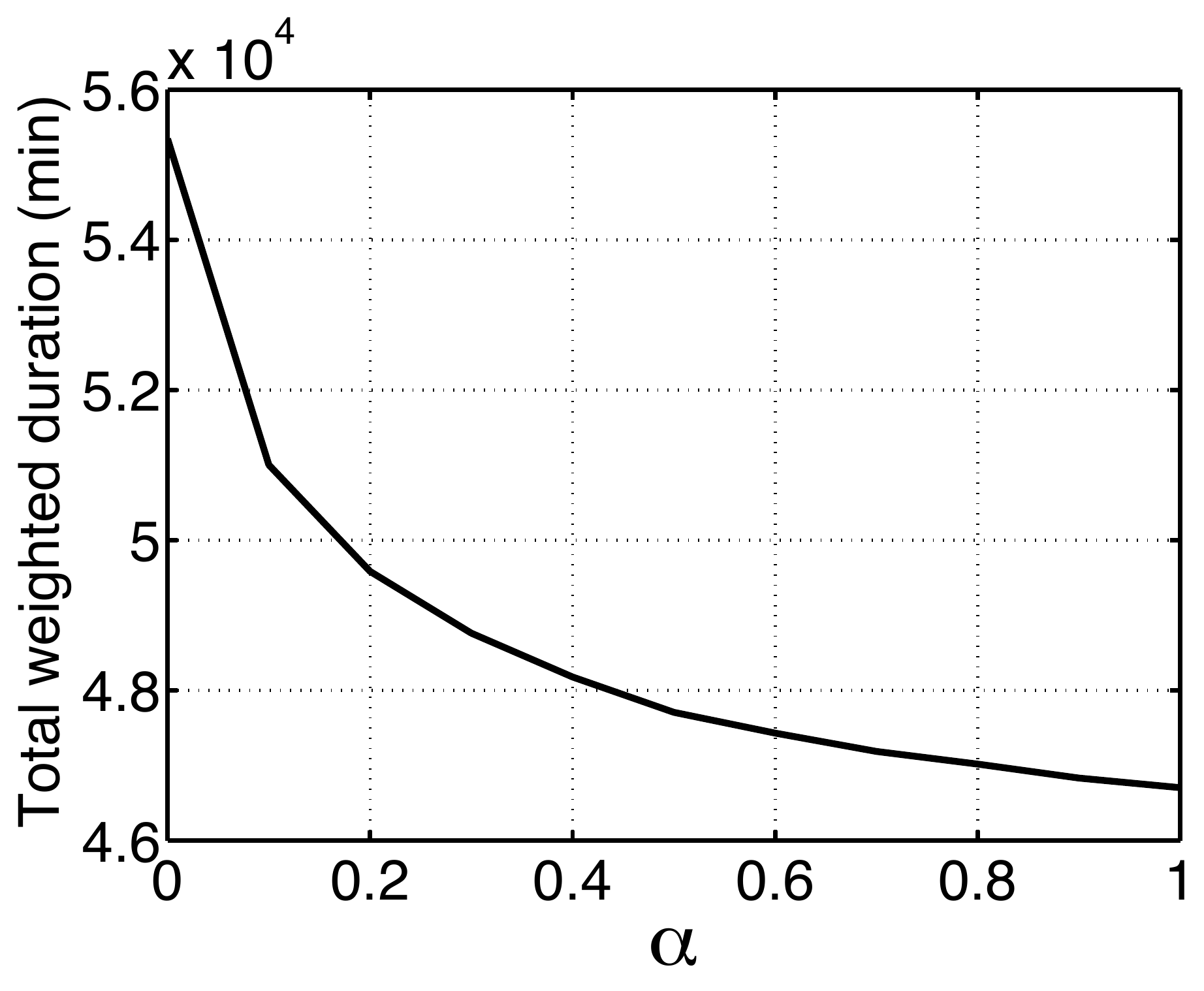}}
 \subfigure[Sum of transit time and weighted gate conflict duration.]{\label{f:tradeoff3}\includegraphics{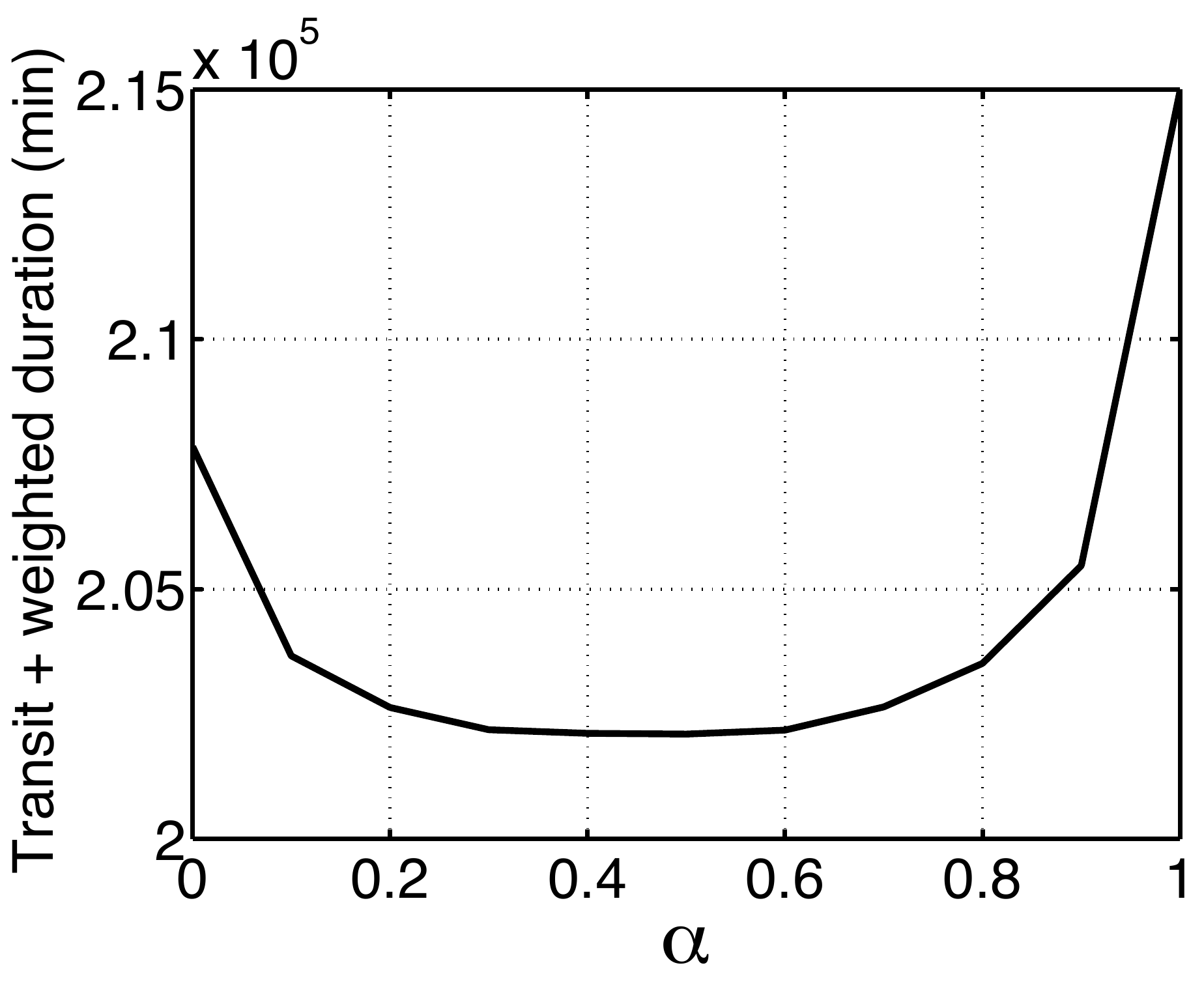}}
\end{subfigmatrix}
\caption{Example 1: $\alpha$ versus total transit time and total weighted gate conflict duration.}
 \label{f:total3}
\end{figure}

Fig.~\ref{f:total3} illustrates changes of passenger transit time and weighted gate conflict duration according to the trade-off factor $\alpha$. As $\alpha$ increases, passenger transit time increases and weighted gate conflict duration decreases. Note that the order of magnitude of passenger transit time is larger than that of weighted gate conflict duration because the duration of gate conflicts is less than few minutes in general as shown in Fig.~\ref{f:duration}. The sum of passenger transit time and weighted gate conflict duration is shown in Fig.~{\ref{f:tradeoff3}}. The sum is minimal when $\alpha$ is in the interval [0.4, 0.5].

\begin{figure}[htb]
 \centering
 \includegraphics[width=0.5\textwidth]{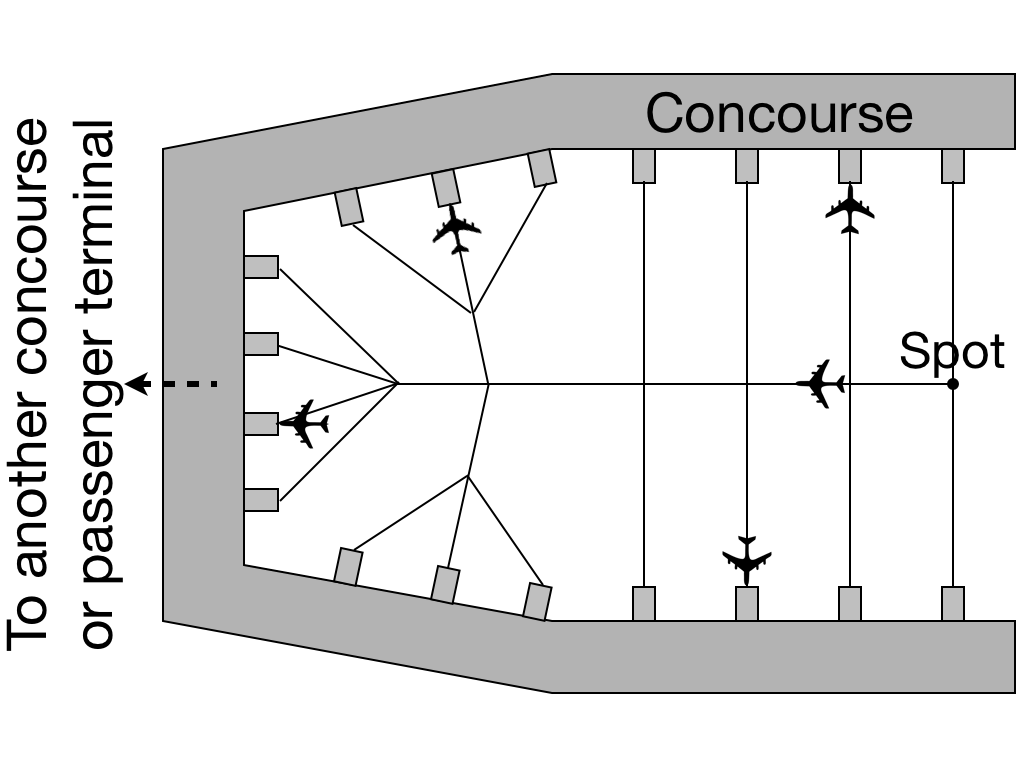}
 \caption{Example 2: Horseshoe configuration}
 \label{f:ramp2}
\end{figure}

\begin{figure}[htb]
\begin{subfigmatrix}{3}
 \subfigure[$\alpha$ versus total transit time.]{\label{f:transit2}\includegraphics{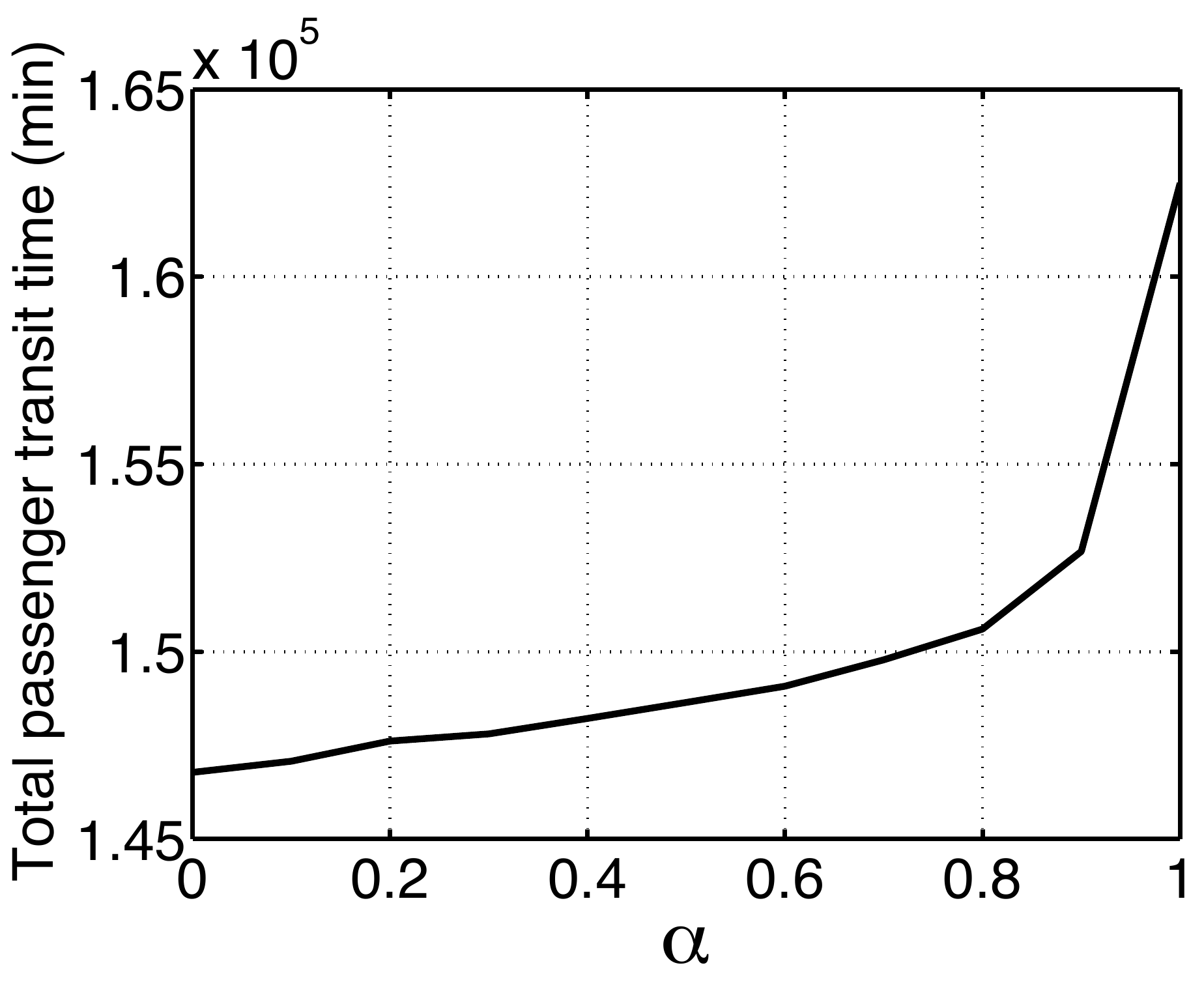}}
 \subfigure[$\alpha$ versus total weighted gate conflict duration.]{\label{f:robust2}\includegraphics{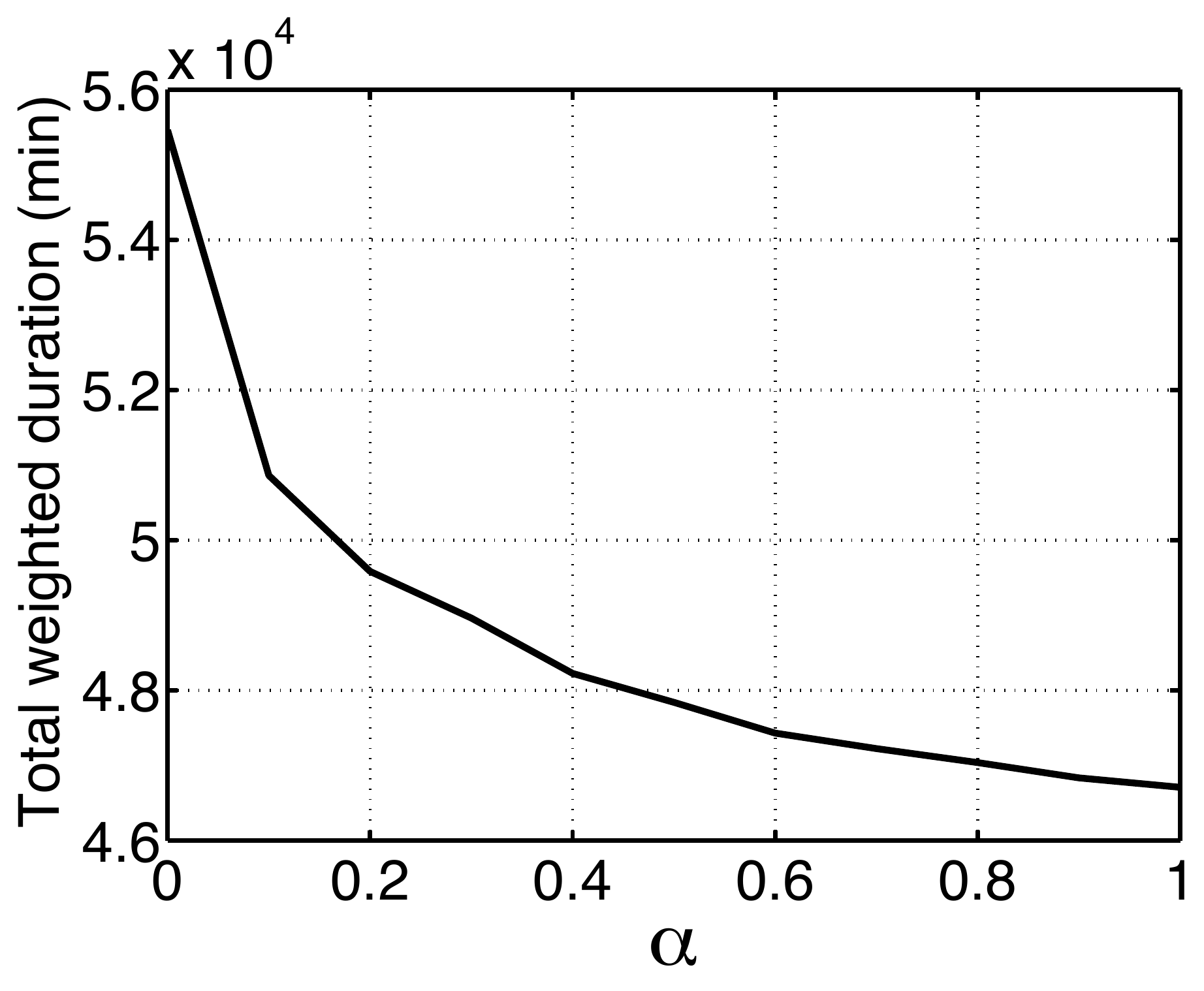}}
 \subfigure[Sum of transit time and weighted gate conflict duration.]{\label{f:tradeoff4}\includegraphics{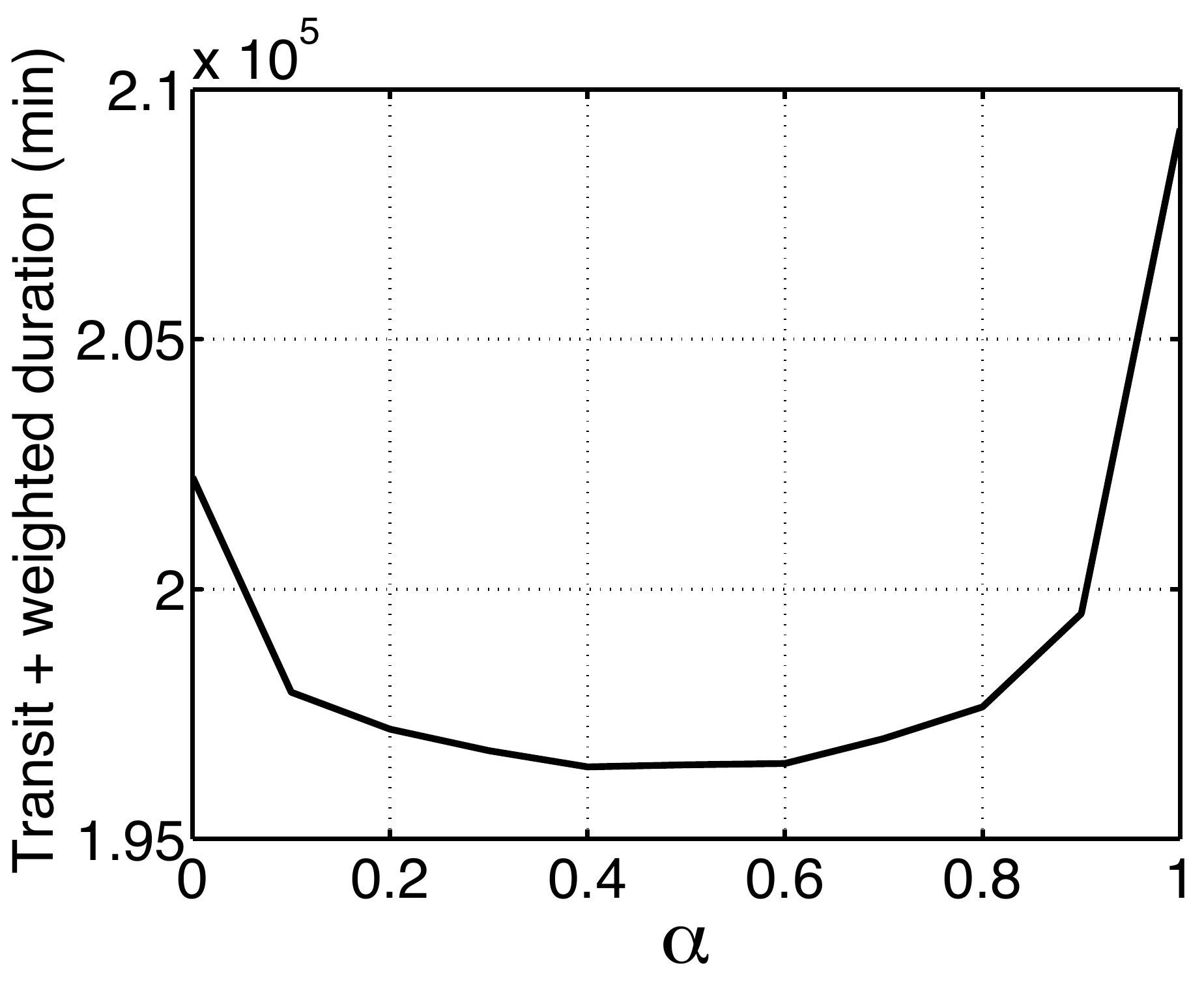}}
\end{subfigmatrix}
\caption{Example 2: $\alpha$ versus total transit time and total weighted gate conflict duration.}
 \label{f:total4}
\end{figure}

The second example that resembles the horseshoe of Boston Logan Airport is shown in Fig.~\ref{f:ramp2}. Passenger transit time, weighted gate conflict duration, and the sum of them are illustrated in Fig.~\ref{f:total4}. They are similar to the example 1. In this case, the sum is minimal when $\alpha$ is equal to 0.4. From two examples, it is concluded that the robust gate assignment reduces gate conflicts but increases passenger transit time, and there is a trade-off between the robustness of gate assignment and passenger transit time.

\section{Conclusion}
In this research, we propose a robust gate assignment problem. We define the robustness of gate assignment and analyze gate delays in order to calculate the expected duration of gate conflicts, which depends on gate separation. Specifically, the expected duration of gate conflicts increases exponentially as gate separation decreases. Simulation shows that the robust gate assignment is more robust than the baseline gate assignment in terms of gate conflict duration and the number of gate conflicts. Moreover, the robust gate assignment is more robust than the baseline gate assignment even when traffic increases. However, the robust gate assignment increases passenger transit time, and there is a trade-off between the robustness of gate assignment and passenger transit time.

\bibliographystyle{model2-names}
\bibliography{rga}

\end{document}